\documentclass[aip,reprint]{revtex4-1}
\usepackage[utf8]{inputenc}

\usepackage{amsmath}
\usepackage{amssymb}
\usepackage{graphicx}
\usepackage[caption=false]{subfig}
\usepackage{xcolor}

\usepackage{longtable}

\draft 

\newcommand{\dt}{\partial_t}
\newcommand{\dg}{\cdot\nabla}
\renewcommand{\div}{\nabla\cdot}
\newcommand{\mi}{m_i}
\newcommand{\md}{m_d}
\newcommand{\mn}{m_n}
\newcommand{\me}{m_e}
\newcommand{\gammai}{\gamma_i}
\newcommand{\gamman}{\gamma_n}
\newcommand{\gammae}{\gamma_e}
\newcommand{\agi}{a_{\gamma i}}
\newcommand{\agn}{a_{\gamma n}}

\renewcommand{\ni}{{n_i}}
\newcommand{\nd}{{n_d}}
\newcommand{\nn}{{n_n}}

\newcommand{\eZ}{\epsilon Z}
\newcommand{\ui}{{u_i}}
\newcommand{\ud}{{u_d}}
\newcommand{\un}{{u_n}}

\newcommand{\cia}{c_{IA}}
\newcommand{\cda}{c_{DA}}
\newcommand{\csi}{c_{si}}
\newcommand{\csn}{c_{sn}}
\newcommand{\cse}{c_{se}}
\newcommand{\nudn}{{\nu_{dn}}}
\newcommand{\nudi}{{\nu_{di}}}
\newcommand{\nund}{{\nu_{nd}}}
\newcommand{\nuni}{{\nu_{ni}}}
\newcommand{\nuid}{{\nu_{id}}}
\newcommand{\nuin}{{\nu_{in}}}
\newcommand{\nuI}{{\nu_I}}
\newcommand{\nuL}{{\nu_L}}

\newcommand{\nuA}{{\nu_A}}
\newcommand{\wsd}{{w_{sd}}}
\newcommand{\wsi}{{w_{si}}}
\newcommand{\tdn}{{t_{dn}}}
\newcommand{\tdA}{{t_{dA}}}
\newcommand{\tdi}{{t_{di}}}
\newcommand{\tni}{{t_{ni}}}
\newcommand{\htdn}{\hat t_{dn}}
\newcommand{\htdi}{\hat t_{di}}
\newcommand{\zsn}{\zeta_{sn}}
\newcommand{\zwn}{\zeta_{wn}}
\newcommand{\zsA}{\zeta_{sA}}
\newcommand{\zwA}{\zeta_{wA}}
\newcommand{\zsi}{\zeta_{si}}
\newcommand{\zwi}{\zeta_{wi}}
\newcommand{\zni}{\zeta_{ni}}

\newcommand{\hwsd}{\hat w_{sd}}

\newcommand{\brhod}{\bar\rho_d}
\newcommand{\Rd}{R_d}
\newcommand{\xI}{x_I}

\begin{document}


\title{Resonant instabilities mediated by drag and electrostatic interactions in laboratory and astrophysical dusty plasmas} 



\author{Ben Y. Israeli}
\email[]{bisraeli@pppl.gov}
\affiliation{Department of Astrophysical Sciences, Princeton University, Princeton, NJ 08544, USA}
\affiliation{Princeton Plasma Physics Laboratory, Princeton, NJ 08540, USA}

\author{Amitava Bhattacharjee}
\affiliation{Department of Astrophysical Sciences, Princeton University, Princeton, NJ 08544, USA}
\affiliation{Princeton Plasma Physics Laboratory, Princeton, NJ 08540, USA}

\author{Hong Qin}
\affiliation{Princeton Plasma Physics Laboratory, Princeton, NJ 08540, USA}
\affiliation{Department of Astrophysical Sciences, Princeton University, Princeton, NJ 08544, USA}


\date{\today}

\begin{abstract}
Dusty plasmas are known to support a diverse range of instabilities, including both generalizations of standard plasma instabilities and ones caused by effects specific to dusty systems. It has been recently demonstrated that a novel broad class of streaming instabilities, termed resonant drag instabilities (RDIs), can be attributed to a particular resonance phenomenon, manifested by defective eigenvalues of the linearized dust/fluid system. In this work, it is demonstrated that this resonance phenomenon is not unique to RDIs and can be used as a framework to understand a wider range of instabilities, termed resonant instabilities. Particular attention is given to the filamentary ionization instability seen in laboratory dusty plasmas and to the two-stream instability. It is shown that, due to the commonalities in underlying physics between the dust-ion-acoustic two-stream instability and the acoustic RDI, these instabilities should be relevant in strongly overlapping regimes in astrophysical dusty plasmas. It is proposed that a similar overlap in the experimental accessibility of these modes (and of the filamentary instability) allows for the possibility of experimental investigation in the laboratory of complex and astrophysically relevant instability dynamics.
\end{abstract}


\maketitle 

\section{Introduction}
The physics of dust particles in fluids or plasmas is a subject of great interest for its potential applications to a wide variety of problems in astrophysics and in laboratory contexts such as soft condensed matter physics and semiconductor manufacturing. Despite shared attributes in the underlying physics, communication between these diverse communities is surprisingly infrequent. 

In one of their several 2018 papers, Squire and Hopkins (hereafter SH) introduce the concept of resonant drag instabilities (RDIs), a general class of instabilities in systems containing dust moving with drag through a fluid which supports oscillatory modes.\cite{squireResonantDragInstability2018} In particular, they demonstrated that virulent instabilities appear when the phase velocity of a wave in the fluid matches the streaming velocity of the dust projected along the wave propagation direction, causing a resonance phenomenon. In subsequent papers, they point out that this framework, and a corresponding perturbative analysis of the resonance, may be used to describe a range of instabilities, many of which are novel.\cite{hopkinsResonantDragInstability2018,hopkinsUbiquitousInstabilitiesDust2018,squireResonantDragInstabilities2018} They also point out that the RDI is analogous to other cases of resonance-induced fluid instabilities, and that their perturbative approach may be applied to general coupled linear systems.

In this paper, we extend their analysis of fluid-dust systems to include forms of interaction other than fluid drag, focusing particularly on electrostatic effects in a dusty plasma in which the dust particles are charged. This was motivated by earlier work by Wang et. al. demonstrating that the filamentary mode seen in laboratory dusty plasma experiments may be treated as the product of a resonance between dust and ion electrostatic modes mediated by drag and ionization dynamics.\cite{wangIonizationInstabilitiesResonant2001} This instability has been attributed to a feedback process, wherein a reduction in dust density in a region corresponds to higher electron density, which in turn increases ionization rate, producing an ion wind which pulls dust away from the region via drag.\cite{wangIonizationInstabilitiesResonant2001,samsonovInstabilitiesDustyPlasma1999}

In Section \ref{sec:instabilities}, we demonstrate that this mode fits within SH's perturbative framework despite occurring at zero drift velocity and requiring deriving the next-order term in the expansion. We further show that the dust-ion-acoustic two-stream instability, involving solely electrostatic interaction, may be similarly modeled, and consider connections to the phenomenon of PT-symmetry breaking. This mode, a particular case of the general two-stream instability seen in plasmas with counter-propagating charged particle populations, drives the amplification of dust-ion-acoustic waves via resonant transfer of energy from streaming dust.\cite{havnesMotionDestructionGrains1980}

In Section \ref{sec:lab}, we compare the growth rates of these two modes with that of the acoustic RDI under the laboratory glow discharge conditions considered by Wang et. al., finding that they should be relevant in overlapping regimes. We also discuss means of producing relevant conditions in the laboratory, particularly the requisite streaming velocities. Levitation of dust in a sheath with large ion velocities, and considerations for electrostatic acceleration of the dust in both ionized and neutral environments are discussed.

In Section \ref{sec:astro}, we then present a similar analysis of several strongly and weakly ionized astrophysical environments. We find that dust-ion-acoustic two-stream instabilities and acoustic RDIs should generally occur simultaneously at sufficient drift velocity, with the dominant mode dependent on wavelength. In particular, we first consider growth rates of both modes in protoplanetary disks, an environment in which the Youdin-Goodman streaming instability\cite{youdinStreamingInstabilitiesProtoplanetary2005}, a streaming instability driven by drag rather than electrostatic forces, is considered to play a significant role in planetesimal formation\cite{youdinStreamingInstabilitiesProtoplanetary2005,johansenProtoplanetaryDiskTurbulence2007,johansenRapidPlanetesimalFormation2007}. (See \citet{squireResonantDragInstabilities2018} for a description of the Youdin-Goodman instability as an "epicyclic" RDI along with several other RDIs of relevance to protoplanetary disks.) We then consider the shocked interstellar medium (ISM), in which the two-stream instability has been studied as a mechanism for dust grain destruction\cite{havnesMotionDestructionGrains1980}, and move on to winds from active galactic nuclei and planetary nebulae, both of which are mentioned by \citet{hopkinsResonantDragInstability2018} as possible locations for the acoustic RDI, and in which we find both the two-stream instability and acoustic RDI to be of relevance.

\section{Model and notation}
\label{sec:model}
We consider instabilities in several related fluid models of dusty plasmas. Here we present the most general case, including all terms utilized below, in order to define the notation and underlying assumptions. We consider a quasi-neutral system consisting of a neutral gas, ions, dust, and electrons. We treat the first three species as fluids, and assume the electron density instantaneously responds adiabatically to the electrical potential.
Our equations are

\begin{widetext}
\begin{equation}
\label{equ:EoM}
\begin{split}
0=&\dt\nd+\div(\ud\nd)\\
0=&\md\nd(\dt\ud+\ud\dg\ud)-\nd eZ\nabla\phi+\md\nd(\nudi(\ud-\ui)+\nudn(\ud-\un))+\mi\ni\nuA(\ud-\ui)-\md\nd a\\ 
0=&\dt\ni+\div(\ui\ni)-\nuI\nn+(\nuL+\nuA)\ni\\
0=&\mi\ni(\dt\ui+\ui\dg\ui)+\ni e\nabla\phi+\mi\csi^2\nabla\ni+\mi\ni(\nuid(\ui-\ud)+\nuin(\ui-\un))+\mi\nn\nuI(\ui-\un)\\ 
0=&\dt\nn+\div(\un\nn)+\nuI\nn\\
0=&\mn\nn(\dt\un+\un\dg\un)+\mn\csn^2\nabla\nn+\mn\nn(\nund(\un-\ud)+\nuni(\un-\ui))\\ 
0=&\ni-n_e-Z\nd\\
0=&\frac{\cse^2\me}{\gammae}\nabla n_e-n_e e\nabla\phi
\end{split}
\end{equation}
\end{widetext}

We assume here the dust to have a mean charge $-Z$ and the ions a mean charge $+1$. For simplicity, variation in dust size, mass, and charge is ignored, with all dust grains taken to be identical with radius $R_d$ and material density $\brhod$. We include interactions between species with rates $\nu$: collisions between species $\nu_{ij}$, ionization of neutrals $\nuI$, adsorption of ions onto dust $\nuA$, and ion loss $\nuL$. Dust-dust collisions are neglected, with dust being treated as a pressure-less fluid. In cases in which ionization processes are included, the ionization fraction is assumed to be small, such that the ionization of species with mean mass $\mi$ has a negligible effect on the mean neutral mass $\mn$.


Equation \ref{equ:EoM} is valid in an inertial frame or, in the case of a uniform gravitational acceleration, a free-falling frame. A dust acceleration term $\md\nd a$ is included to support an arbitrary equilibrium dust streaming velocity (relative to the neutrals) $\wsd=u_{d0}-u_{n0}$  that is non-vanishing and constant. In general, this will also induce a nonzero equilibrium ion streaming velocity $\wsi$.  The linearization of equation \ref{equ:EoM} is performed in a frame accelerating with the system, such that the Fourier ansatz used for our first-order perturbations is
\begin{equation}
    \delta n, \delta u... \propto \exp{\left(i(k\cdot x-(\omega+\frac12 k\cdot a_{eff} t)t)\right)}
\end{equation}
where $a_{eff}$ is the net acceleration of the total system. This frame is chosen such that equilibrium flow velocities are all constant. See Appendix B of Hopkins and Squire\cite{hopkinsResonantDragInstability2018} for the definition of a similar frame.\footnote{Hopkins and Squire\cite{hopkinsResonantDragInstability2018} utilize a frame co-moving with the gas, whereas our calculations were performed in a frame with a constant nonzero equilibrium neutral velocity. These frames are related by a Galilean transformation, and the distinction has no impact on the linearized system.} We will work with the dust streaming velocity as a tunable parameter which determines the required acceleration $a$ and resulting ion streaming velocity.

If our equations seem unduly complex, it is because we aspire to treat both astrophysical and laboratory applications from a common perspective. A list of symbols used throughout the paper is given in Appendix \ref{sec:symbols}, and parameter values for each model considered below are listed in Appendix \ref{sec:values}.

\subsection{Wave speeds}
We will frequently use the sound velocity of each species, the dust-acoustic velocity, and the dust-ion-acoustic velocity, which we define (at equilibrium) as follows:
\begin{equation}
\begin{split}
c^2_{sj}=&\frac{\gamma_j T_j}{m_j}\\
\cda^2=&\frac{\epsilon\me Z^2\cse^2}{\gammae\md(1-\epsilon Z)}\\
\cia^2=&\frac{\me\cse^2}{\gammae\mi(1-\eZ)}
\end{split}
\label{equ:speeds}
\end{equation}
where $\gamma_j$ is the polytropic index for species $j$.

\subsection{Collision rates}
\label{sec:collisions}
Per Hopkins and Squire, we treat collision rates in terms of a collision time: $\nudn=\frac1\tdn,\ \nudi=\frac1\tdi,\ \nuA=\frac1\tdA,\ \nuni=\frac1\tni$. Due to momentum conservation, $\nund,\ \nuid,\ \nuin$ are determined by
\begin{equation}
\begin{split}
    \mn\nn\nund=&\md\nd\nudn\\
    \mi\ni\nuid=&\md\nd\nudi\\
    \mi\ni\nuin=&\mn\nn\nuni
\end{split}
\end{equation}

In the process of linearization, we must consider the first order perturbations to collision rates due to their dependence on the density and velocity of each species. Mimicking the notation of Hopkins and Squire\cite{hopkinsResonantDragInstability2018}, perturbations are linearized as
\begin{equation}
\begin{split}
\tdn=&\tdn_0\left(1-\zsn\frac{\nn_1}{\nn_0}-\zwn\frac{(\ud_0-\un_0)\cdot(\ud_1-\un_1)}{|\ud_0-\un_0|^2}\right)\\
\tdi=&\tdi_0\left(1-\zsi\frac{\ni_1}{\ni_0}-\zwi\frac{(\ud_0-\ui_0)\cdot(\ud_1-\ui_1)}{|\ud_0-\ui_0|^2}\right)\\
\tdA=&\tdA_0\left(1-\zsA\frac{\ni_1}{\ni_0}-\zwA\frac{(\ud_0-\ui_0)\cdot(\ud_1-\ui_1)}{|\ud_0-\ui_0|^2}\right)\\
\tni=&\tni_0\left(1-\zni\frac{\ni_1}{\ni_0}\right)\\
\end{split}
\label{equ:colcoeffs}
\end{equation}

The dimensionless coefficients $\zeta$ are determined by the drag physics. Their values, along with expressions for the collision times $t$, for cases of interest in this paper are given in Appendix \ref{sec:collision-physics}.

\subsection{Dimensionless quantities}

For expedience of notation, we use the following definitions, partially borrowed from Hopkins and Squire, and Wang et. al.\cite{hopkinsResonantDragInstability2018,wangIonizationInstabilitiesResonant2001}

\begin{equation}
\begin{split}
\epsilon=&\nd/\ni\\
\alpha=&(1-\eZ)^{-1}\\
\mu=&\frac{\md\nd}{\mn\nn}\\
\xI=&\frac{\ni}{\nn}\\
\cos\theta=&\frac{\mathbf{k}\cdot\mathbf{w}_{sd}}{|\mathbf{k}||\mathbf{w}_{sd}|}
\end{split}
\label{equ:dimless}
\end{equation}

\section{Resonant instabilities}
\label{sec:instabilities}
In this section, we extend the defective eigenvalue approach put forward by SH to describe RDIs\cite{squireResonantDragInstability2018} to other dusty plasma instabilities. As such, we begin by reviewing their derivation of the growth rate of the RDI, and then perform analogous derivations for the other instabilities. The reason for the general applicability of this approach to instabilities in dusty plasmas is that the extremely large difference between dust and ion/neutral masses, and the generally substantial difference between dust and ion/neutral densities, provide natural parameters with which to perform a perturbative analysis.

\subsection{Review of resonant drag instabilities}
\label{sec:RDI}
As demonstrated by SH, RDIs appear in any system containing dust streaming through a fluid which supports at least one oscillatory mode. Their growth rates tend to peak dramatically under conditions in which the phase velocity of the oscillation matches the streaming velocity of the dust projected along the wave propagation direction. The wave then appears static in the dust's frame, allowing feedback between dust and fluid perturbations. Their general description of such instabilities has been shown to be applicable to understanding previously modeled modes such as the Youdin-Goodman streaming instability\cite{youdinStreamingInstabilitiesProtoplanetary2005}, as well as novel modes such as acoustic and magnetosonic RDIs.\cite{squireResonantDragInstability2018,hopkinsResonantDragInstability2018,hopkinsUbiquitousInstabilitiesDust2018,squireResonantDragInstabilities2018}

Before discussing generalization to electrostatic systems, we review the matrix formalism of SH, following their notation conventions. In circumstances in which a small parameter (generally the equilibrium dust-gas mass density ratio $\mu$) allows the action of the dust on the fluid to be treated perturbatively, the growth rate of the RDI at resonance can be calculated as the perturbation of a defective eigenvalue of a matrix.

Suppose that the linearization of the fluid equations describing our dust-fluid system is given by
\begin{equation}
    i\dot{\vec x}=\omega\vec x=T\vec x=(T_0+\mu T_1)\vec x
\label{equ:linear}
\end{equation}
for some state vector $\vec x=(\delta n_d,\delta \mathbf{u}_d,\vec x_F)$ of the dust density, dust velocity, and fluid state variables (generally density and velocity for one or more species), with $T_0$ and $T_1$  being block matrices.

\begin{equation}
    T_0=
    \left(\begin{array}{cc}
            A & C \\
            0 & F
    \end{array}\right),\ \ 
    T_1=
    \left(\begin{array}{cc}
        T_1^{AA} & T_1^{AF} \\
        T_1^{FA} & T_1^{FF}
    \end{array}\right)
    \label{equ:T0T1}
\end{equation}

The diagonal blocks correspond to the dynamics of the dust ($A$, $T_1^{AA}$) and fluid ($F$, $T_1^{FF}$) independent of each other at zeroth and first order in $\mu$. The off-diagonal blocks describe the effect of the fluid on the dust ($C$, $T_1^{AF}$) at zeroth and first order in $\mu$, and the dust on the fluid ($T_1^{FA}$) at first order in $\mu$. If $A$ and $F$ share an eigenvalue $\omega_0$(assumed to be nondegenerate for simplicity), then $T_0$ has eigenvalue $\omega_0$ with degeneracy 2 but only one eigenvector for $\omega_0$; the matrix is defective. Unlike in non-defective perturbation theory, in which the perturbation of eigenvalues grows linearly, in this case the perturbation grows as $\mu^{1/2}$.
\begin{equation}
    \label{equ:RDIgrowth}
    \omega_1=\pm\mu^{1/2}\left((\xi_F^LT_1^{FA}\xi_A^R)(\xi_A^LC\xi_F^R)\right)^{1/2}
\end{equation}
where $\xi_A^R,\ \xi_A^L,\ \xi_F^R,\ \xi_F^L$, are right and left eigenvectors of $A$ and $F$ for eigenvalue $\omega_0$ normalized such that $\xi_A^L\xi_A^R=\xi_F^L\xi_F^R=1$. In Section \ref{sec:filamentary-resonance}, we provide a novel formula for the next-order correction.

For the typical case of neutral streaming dust in which
\begin{gather*}
    A=
    \left(\begin{array}{cc}
        k\cdot w_s & k^T \\
        0 & k\cdot w_s I+D
    \end{array}\right),\ \ 
    C=
    \left(\begin{array}{c}
        0  \\
        C_\nu 
    \end{array}\right),\ \ 
    \omega_0=k\cdot w_s
\end{gather*}
where $D$ and $C_\nu$ are block matrices containing drag terms dependent on the details of the fluid model.
This yields
\begin{equation}
    \omega\approx k\cdot w_s\pm i\mu^{1/2}\left((\xi_F^LT_1^{(1)})(k^TD^{-1}C_\nu\xi_F^R)\right)^{1/2}
\end{equation}
$T_1^{(1)}$ is the left column of $T_1^{FA}$. In reality, the instability will still exist away from this resonance or with $w_s$ below the wave velocity $c_s$. Its maximum growth rate will occur near
\begin{equation}
    \theta=\begin{cases}
    \arccos(c_s/w_s) & w_s\ge c_s\\
    0 & w_s<c_s
    \end{cases}
\end{equation}
which we will call the "resonant angle".

\subsection{Two-stream Instabilities}
\label{sec:2S}
Two-stream instabilities appear in a wide range of fluid and kinetic systems in which two charged particle populations move relative to one another. Various versions of this instability involving charged dust grains streaming in a plasma have been studied in both fluid and kinetic models.\cite{bharuthramTwostreamInstabilitiesUnmagnetized1992,havnesMotionDestructionGrains1980,havnesStreamingInstabilityInteraction1988,rosenbergIonDustacousticInstabilities1993,rosenbergIonDustTwostream2004,rosenbergModifiedTwostreamInstabilities1995} Here we consider the particular case of dust, which is assumed to be without velocity dispersion, streaming through a warm ion fluid, focusing on conditions in which the dust-ion-acoustic (DIA) mode is driven unstable. It is worth noting that the dust acoustic (DA) mode may also be driven unstable\cite{rosenbergIonDustacousticInstabilities1993}, but as this mode is not amenable to our perturbative framework, we do not consider it here. Throughout this paper, the term two-stream instability will be used to refer to that of the DIA mode. The equations are

\begin{equation}
\begin{split}
0=&\dt\nd+\div(\ud\nd)\\
0=&\md\nd(\dt\ud+\ud\dg\ud)-\nd eZ\nabla\phi\\
0=&\dt\ni+\div(\ui\ni)\\
0=&\mi\ni(\dt\ui+\ui\dg\ui)+\ni e\nabla\phi+\mi\csi^2\nabla\ni\\
0=&\ni-n_e-Z\nd\\
0=&\frac{\cse^2\me}{\gammae}\nabla n_e-n_e e\nabla\phi
\end{split}
\label{equ:2s}
\end{equation}

Since the dust grains are generally much more massive than the ions, we may consider the action of the ions on the dust small (inverse to the normal RDI case). Linearizing and taking $\mi/\md$ to be a small parameter we arrive at $\omega\vec x=(T_0+\frac\mi\md T_1)\vec x$ with
\begin{equation}
\label{equ:2Smat}
\begin{split}
T_0=&\left(
\begin{array}{cccc}
\mathbf{k}\cdot \mathbf{w}_{sd} & \mathbf{k}^T & 0 & 0\\
0 & \mathbf{k}\cdot \mathbf{w}_{sd} \mathbf{I} & 0 & 0\\
0 & 0 & 0 & \mathbf{k}^T\\
-\eZ \cia^2 \mathbf{k} & 0 & (\cia^2+\csi^2)\mathbf{k} & 0
\end{array}
\right)\\
T_1=&\left(
\begin{array}{cccc}
    0 & 0 & 0 & 0\\
    \epsilon Z^2 \cia^2 \mathbf{k} & 0 & -Z\cia^2 \mathbf{k} & 0\\
    0 & 0 & 0 & 0\\
    0 & 0 & 0 & 0\\
\end{array}
\right)
\end{split}
\end{equation}

In this case the dominant species is the dust rather than the ions. Taking the transpose of these matrices would make them match the form of equation \ref{equ:T0T1}. The dust and ion sectors of $T_0$ share an eigenvalue, indicating a resonance between the dust streaming velocity and the DIA wave, for $\omega_0=\mathbf{k}\cdot \mathbf{w}_{sd}=\pm|\mathbf{k}|\sqrt{\cia^2+\csi^2}$, or equivalently for $\wsd\cos(\theta)=\pm\sqrt{\cia^2+\csi^2}$. Note that this eigenvalue is triply defective since the dust sector is itself defective. Calculating the perturbation to lowest order in $\frac\mi\md$, we find that for both the positive and negative cases the growth rate of the unstable branch is
\begin{equation}
    \Im(\omega)\approx \frac{\sqrt{3}(\epsilon\frac\mi\md)^{1/3}(k^2|Z|\cia^2)^{2/3}}{2^{4/3}(\mathbf{k}\cdot \mathbf{w}_{sd})^{1/3}}=\frac{\sqrt{3}(k^2\cda\cia)^{2/3}}{2^{4/3}(\mathbf{k}\cdot \mathbf{w}_{sd})^{1/3}}
\label{equ:2Sgrowth}
\end{equation}
Using the parameters of Section \ref{sec:Wang-model}, this estimate was found numerically to have an apparently constant deviation from the exact result of $0.45\%$ for a range of frequencies and wavelengths.

\begin{figure}
\subfloat[]{
\includegraphics[width=0.3\textwidth]{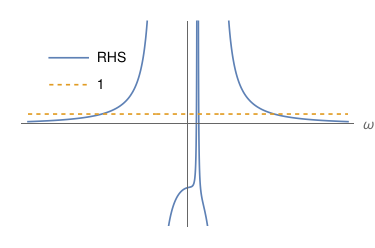}
}

\subfloat[]{
\includegraphics[width=0.3\textwidth]{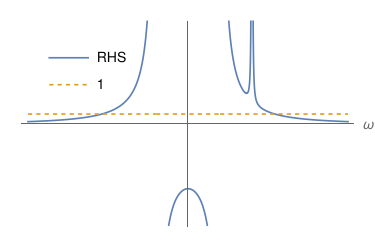}
}

\subfloat[]{
\includegraphics[width=0.3\textwidth]{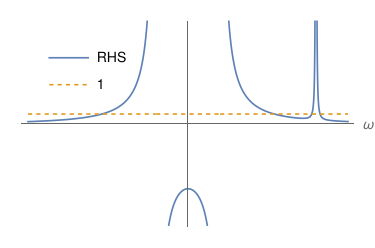}
}

\caption{The left and right-hand sides of equation \ref{equ:twostream} (plotted as separate functions) for $\mathbf{k}||\mathbf{w}_{sd}$ with increasing streaming velocity, plotted as a function of $\omega$. For (a) low streaming velocity  or (c) high streaming velocity, there are four real roots, indicated by the intersections of the plotted functions. However, for (b) intermediate streaming velocity, there may be only two real roots, indicating the existence of a pair of complex roots.}
\label{fig:two-stream-demo}
\end{figure}

Instead of the perturbative approach, we may of course also solve for the exact dispersion relation, yielding
\begin{equation}
    1=\frac{|\mathbf{k}|^2\cia^2}{\omega^2-|\mathbf{k}|^2\csi^2}+\frac{|\mathbf{k}|^2\cda^2}{(\omega-\mathbf{k}\cdot\mathbf{w}_{sd})^2}
    \label{equ:twostream}
\end{equation}
In the limit $\cda\ll\cia,\ \csi$, it is easily shown that this has a pair of complex roots for $\csi<\frac{|\mathbf{k}\cdot\mathbf{w}_{sd}|}{|\mathbf{k}|}=\wsd|\cos\theta|<\sqrt{\cia^2+\csi^2}$ by considering the relative location of the singularity in the second term on the right hand side. This is illustrated in Figure \ref{fig:two-stream-demo}.

\subsubsection{Remark on Resonance and PT-symmetry}
\begin{figure}
    \subfloat[]{
    \includegraphics[width=0.45\textwidth]{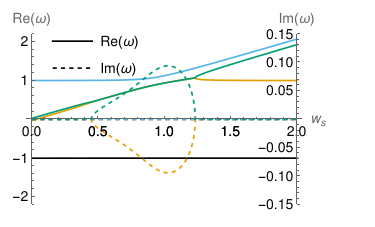}
    }
    
    \subfloat[]{
    \includegraphics[width=0.45\textwidth]{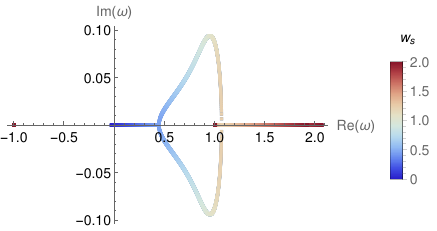}
    }
    
    \caption{Eigenvalues of the two-stream matrix defined in equation \ref{equ:2Smat} assuming parallel propagation, in units normalized to $k$ and $\sqrt{\cia^2+\csi^2}$, with $\mi/\md=1/200$, $\epsilon=1$, $Z=1$, $\csi/\cia=1/2$. $\mi/\md$ is taken to be comparably large in order to more clearly illustrate distinct eigenvalues. Real and imaginary components plotted a) as a function of streaming velocity b) in the complex plane for varied streaming velocity. The negative real-valued mode is an ion-acoustic wave propagating opposite to streaming.}
    \label{fig:PT-2S}
\end{figure}

It is worth noting that the source of the $(\mi/\md)^{1/3}$ scaling in equation \ref{equ:2Sgrowth} (as opposed to the $(\mi/\md)^{1/2}$ scaling that might be expected from comparison to equation \ref{equ:RDIgrowth}) is that the resonant eigenvalue has a degeneracy of three rather than two due to a resonance within the dust sector.\cite{squireResonantDragInstability2018} This resonance becomes exact to all orders at the boundary of the unstable region of parameter space. The system is PT-symmetric (as apparent in all matrix elements being real), such that its eigenvalues are symmetric with respect to the real axis, and can only leave and enter the real axis as complex conjugate pairs, requiring a resonance at the point of instability onset, known as a Krein resonance.\cite{benderMakingSenseNonHermitian2007,qinKelvinHelmholtzInstabilityResult2019,zhangStructureTwostreamInstability2016} This is illustrated in Figure \ref{fig:PT-2S}. As such, the instability is associated with the spontaneous breaking of PT symmetry, in the sense that the conjugate eigenvectors are not PT-symmetric.\cite{benderMakingSenseNonHermitian2007,qinKelvinHelmholtzInstabilityResult2019} See Appendix \ref{sec:PT} for a review of the role of PT-symmetry in the stability of classical systems, as well as a consideration of possible relevance to RDIs.

\subsection{Filamentary ionization instability}
\label{sec:filament}
Several experiments in laboratory dusty plasmas have observed a rapidly growing instability that results in the formation of filamentary structures in the dust density and glow.\cite{praburamExperimentalObservationVery1996,samsonovInstabilitiesDustyPlasma1999} Wang et. al. investigated a model of this instability in which it is precipitated by a resonance between dust-ion-acoustic and dust-acoustic modes.\cite{wangIonizationInstabilitiesResonant2001} Here, we demonstrate that this instability can be treated as a resonant instability that occurs without streaming, caused by both drag and ionization dynamics.

\subsubsection{Model}
\label{sec:Wang-model}
To match our equations to those used by Wang et. al, we assume that there is no adsorption of ions onto dust ($\nuA=0$), and that ionization of neutrals is driven by collisions with electrons, resulting in $\nuI=\frac{n_e}\nn \nu_{Ie}$, where $\nu_{Ie}$ is a constant. The ion loss rate $\nuL$ is taken to be a constant. Note that in the zero streaming velocity case considered by Wang et. al, perturbations to the collision rates are irrelevant at first order since the velocities they are multiplied by in the momentum equations are first order. For the nozero streaming velocity case, we assume the ion drag on the dust is Coulumb drag and the neutral drag on the dust is Epstein drag. We take the neutrals to be a static uniform background and therefore drop the neutral continuity and momentum equations. Note that Wang et. al. include an additional term ($-\nuL\ni\mi\ui$) in the ion momentum equation. We will replicate their dispersion relation numerically with that term included and will demonstrate that it does not qualitatively affect their results. When comparing against Wang et. al.'s results, the symbol $A_W$ will be used to allow us to simultaneously consider including ($A_W=1$) or not including ($A_W=0$) this extra term. Otherwise, $A_W$ will be assumed to be zero. The ion momentum equation is

\begin{equation}
\begin{split}
0=&\mi\ni(\dt\ui+\ui\dg\ui)+\ni e\nabla\phi+\mi\csi^2\nabla\ni\\
&-A_W\nuL\ni\mi\ui+\mi\ni(\nuid(\ui-\ud)\\
&+\nuin(\ui-\un))+\mi\nn\nuI(\ui-\un)\\
\end{split}
\end{equation}

Wang et. al. use the following values, relevant to laboratory experiments: $T_e=3$eV, $T_i(=T_n)=0.025$eV, $\ni_0=2\times10^9$cm$^{-3}$, $\nd_0=10^7$cm$^{-3}$, $\nn_0=1.33\times10^{16}$cm$^{-3}$, $Z=75$, $\nuin=10^6$s$^{-1}$, $\nudn=5.6\times10^3$s$^{-1}$, $\nuid=3.47\times10^5$s$^{-1}$, $\nuL=5\times10^5$s$^{-1}$, $\csi=3.5\times10^4$cm/s, $\mi=\mn=40m_p$ (argon), $\md=8\times10^8m_p$, $\gammae=1$.

\begin{figure}
\subfloat[]{
\includegraphics[width=0.45\textwidth]{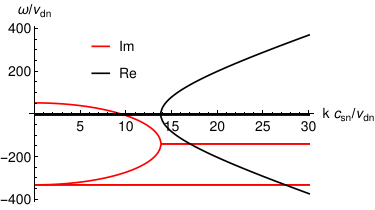}
}

\subfloat[]{
\includegraphics[width=0.45\textwidth]{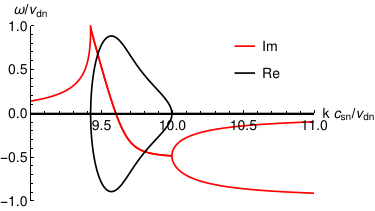}
}

\subfloat[]{
\includegraphics[width=0.45\textwidth]{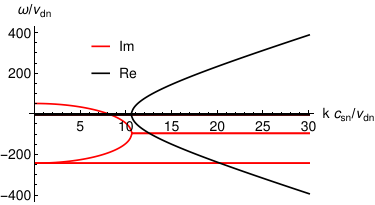}
}

\subfloat[]{
\includegraphics[width=0.45\textwidth]{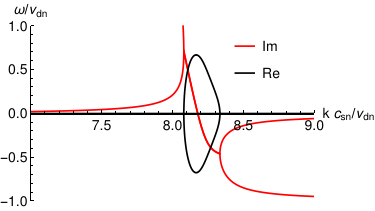}
}
\caption{Dispersion relations for Wang et. al.'s system\cite{wangIonizationInstabilitiesResonant2001} with (a) $A_W=0$ and (c) $A_W=1$, cropped to highlight the resonance with (b) $A_W=0$ and (d) $A_W=1$}
\label{fig:Wang}
\end{figure}

Dispersion relations for this system, with $A_W=0$ and $A_W=1$, are plotted in figure \ref{fig:Wang}.

\subsubsection{Resonant instability}
\label{sec:filamentary-resonance}
Here we demonstrate that the filamentary instability can be framed in the matrix formalism of SH.

Taking zeroth order velocities to be zero, linearizing the equations, and reducing them to a matrix, we expand in $\mi/\epsilon\md=10^{-5}$, yielding $T=T_0+\frac\mi{\epsilon\md} T_1$ with
\begin{widetext}
\begin{equation}
T_0=\left(
\begin{array}{cccc}
 0 & c_{{sn}} \mathbf{k}^T & 0 & 0 \\
 0 & -i \nu _{{dn}} \mathbf{I} & 0 & 0 \\
 -i (\alpha -1) \nu _{{L}} & 0 & i (\alpha -1) \nu _{{L}} & c_{{sn}} \mathbf{k}^T \\
 -\frac{\epsilon Z c_{{IA}}^2}{c_{{sn}}} \mathbf{k} & i \nu _{{id}} & \frac{ \left(c_{{IA}}^2+c_{{si}}^2\right)}{c_{{sn}}} \mathbf{k} & -i \left(\left(1-A_W\right) \nu _{{L}}+\nu _{{id}}+\nu _{{in}}\right)\mathbf{I} \\
\end{array}
\right),\ \ \ 
T_1=\left(
\begin{array}{cccc}
 0 & 0 & 0 & 0 \\
 \frac{\epsilon^2 Z^2 c_{{IA}}^2}{c_{{sn}}}\mathbf{k} & - i \nu _{{id}}\mathbf{I} & -\frac{\epsilon Z c_{{IA}}^2}{c_{{sn}}} \mathbf{k}& i \nu _{{id}} \mathbf{I}\\
 0 & 0 & 0 & 0 \\
 0 & 0 & 0 & 0 \\
\end{array}
\right)
\end{equation}
\end{widetext}
Computing numerical values for the matrix elements, it is clear that all of the nonzero elements of $\frac\mi{\epsilon\md} T_1$ are small compared to those of $T_0$ for $|\mathbf k|\lesssim30cm^{-1}$.\\

As in the case of the two-stream instability and unlike the standard RDI treatment, the dust affects the fluid but the fluid does not affect the dust at lowest order. We note that the dust sector has a zero eigenvalue (at lowest order), and that the ion sector has a zero eigenvalue under the condition:
\begin{equation}
    (\cia^2+\csi^2)\mathbf{k}^2=(\alpha-1)\nuL(\nuid+\nuin+(1-A_W)\nuL)
    \label{equ:resonanceWang}
\end{equation}

For $A_W=1$, this is precisely Wang et. al.'s resonance condition. We therefore expect to be able to treat their instability as analogous to an RDI and calculate its growth rate perturbatively. The relevant modes are longitudinal, so we drop to one dimension. We also take $T_0\rightarrow T_0^T,\ T_1\rightarrow T_1
^T$, in order to match the upper triangular form of SH.

Recall that:
\begin{equation}
    \omega_1=\pm\mu^{1/2}\left((\xi_F^LT_1^{FA}\xi_A^R)(\xi_A^LC\xi_F^R)\right)^{1/2}
\end{equation}

Substituting in numbers and solving equation \ref{equ:resonanceWang} for resonant wavenumber $k_r$, this gives
\begin{equation}
\begin{split}
    k_r=&
    \begin{cases}
    2.18cm^{-1} & A_W=0\\
    1.87cm^{-1} & A_W=1
    \end{cases}\\
    \omega_1=&
    \begin{cases}
    \pm6390s^{-1} & A_W=0\\
    \pm4460s^{-1} & A_W=1 
    \end{cases}\\
\end{split}
\end{equation}

For $A_W=1$, $k_r$ matches Wang et. al.'s value. The values of $\omega_1$ seem to approximate the real part of the frequency in the vicinity of the resonance but do not have an imaginary component. We can recover an approximate imaginary component by calculating the next term in the perturbatative expansion. This can be shown to be
\begin{equation}
\begin{split}
\omega_2=&\frac12\mu(\xi_A^LC\xi_F^R\left(\xi_F^LT_1^{FA}\chi^R+\chi^LT_1^{FA}\xi_A^R\right)\\
&+\xi_A^LT_1^{AA}\xi_A^R+\xi_F^LT_1^{FF}\xi_F^R)
\end{split}
\end{equation}
where $\chi^R$ and $\chi^L$ are the solutions to
\begin{equation}
\begin{split}
    (A-I\omega_0)\chi^R=&\xi_F^L-\frac{C\xi_F^R}{\xi_A^LC\xi_F^R}\\
    \chi^L(F-I\omega_0)=&\xi_F^L-\frac{\xi_A^LC}{\xi_A^LC\xi_F^R}
\end{split}
\end{equation}
such that $\chi^L\chi^R=0$. This evaluates to
\begin{equation}
\begin{split}
    \omega_2=&\begin{cases}
    3620i\ s^{-1} & A_W=0\\ 
    1750i\ s^{-1} & A_W=1 
    \end{cases}\\
\end{split}
\end{equation}
For comparison, the exact value for the frequency of this unstable mode at the resonant wavelength (computed by solving for the eigenvalues of the full matrix) is
\begin{equation}
\begin{split}
    \omega=&\begin{cases}
    \pm4920+1510i\ s^{-1} & A_W=0\\ 
    \pm3720+965i\ s^{-1} & A_W=1
    \end{cases}\\
\end{split}
\end{equation}

\subsubsection{Nonzero streaming velocity}
For nonzero dust streaming velocity, assuming the ion streaming velocity is negligible, the relevant eigenvalue is now $\omega_0=\mathbf{k}\cdot\mathbf{w}_{sd}$, meaning we are seeking a resonance between the dust drift and a fluid wave, as in the normal RDI treatment. The ion sector has a resonant eigenvalue for
\begin{widetext}
\begin{equation}
\left|
\begin{array}{cc}
 i (\alpha -1) \nu _{{L}}-\mathbf{k}\cdot\mathbf{w}_{sd} & c_{{sn}} \mathbf{k}^T \\
 \frac{ \left(c_{{IA}}^2+c_{{si}}^2\right)}{c_{{sn}}} \mathbf{k}-i \frac{(1-\zsi)\nuid_0}{\csn} \mathbf{w}_{sd} & -i \left(\left(1-A_W\right) \nu _{{L}}+\nu _{{in}}\right)\mathbf{I}+\mathbf{D}_i-\mathbf{k}\cdot\mathbf{w}_{sd}\mathbf{I} \\
\end{array}
\right|=0
\end{equation}
\end{widetext}
where
\begin{equation}
\mathbf{D}_i=-i\nuid\left(
\begin{array}{ccc}
     1 & 0 & 0 \\
     0 & 1 & 0 \\
     0 & 0 & 1+\zeta_{wi}
\end{array}
\right)
\end{equation}

\begin{figure}
    \includegraphics[width=0.5\textwidth]{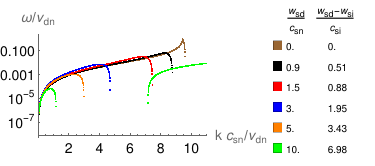}
    \caption{Growth rate of modes coupled to dust in the laboratory system, parallel to streaming, with varied dust streaming velocity. The ionization instability is suppressed as streaming velocity increases. Modes coupled to dust were selected by only plotting eigenvalues whose (unit) eigenvector's dust density and dust velocity components have a combined norm-squared $>5\times10^{-6}$. The two-stream instability is also visible as an additional mode appearing at high streaming velocity.}
    \label{fig:Wang-streaming}
\end{figure}

It can be shown that for parallel propagation, two solutions for $k$ pick up an imaginary component proportional to $\wsd$, while two are purely imaginary. For perpendicular propagation, the zero-drift resonance condition, equation 
\ref{equ:resonanceWang}, is recovered. We therefore expect the instability to be suppressed parallel to a nonzero streaming velocity and to be unchanged perpendicular to streaming. This suppression is illustrated in Figure \ref{fig:Wang-streaming}.

\subsubsection{Ionization in astrophysical settings}
\label{sec:astroion}
The previously discussed filamentary instability relies on the feedback of ion density on the ionization process via its electrostatic modulation of electron density. With this feedback removed, the instability vanishes. The appearance of an RDI-like instability involving ionization processes motivates considering whether related pathways might induce resonant instabilities in other systems. In astrophysical weakly ionized dusty plasmas, such as in protoplanetary disks or the ISM, the ionization fraction is often determined by balance between ionization of neutrals by cosmic rays (taken as a constant background rate) and adsorption of ions onto dust grains. With several RDIs already theorized to drive dust density fluctuations in these settings, the variation of adsorption rate with dust density might be considered as a feedback mechanism to drive a novel resonant instability without dust drift. Here we demonstrate that no such instabilities appear.

\paragraph{Model}
We consider a weakly ionized gas containing dust which is uniformly accelerated relative to the neutrals and ions, inducing a nonzero streaming velocity of both the dust and ions (via drag interactions) relative to the neutrals. We will assume that all ions colliding with dust are adsorbed ($\nudi=\nuid=0$). This is valid if the electrostatic potential at the dust surface is negligible compared to the ion kinetic energy i.e. $T_i\gg Ze^2/\Rd$. Taking all species to be at the same temperature, this is parameterized by $\Gamma=\frac{Ze^2}{\Rd T}$, which we take to be small. Ionization is assumed to be dominated by external radiation sources, giving a constant ionization rate $\nuI$. Ion loss is dominated by adsorption ($\nuL=0$).

\paragraph{Analysis}
With zero dust streaming velocity, we have a similar story to Section \ref{sec:filamentary-resonance}. We take the ion and dust sectors for now, treating the neutrals as a static background. Expanding in $\mi/\epsilon\md$, our zeroth order matrix is
\begin{equation}
T_0=\left(
\begin{array}{cccc}
 0 & c_{{sn}} \mathbf{k}^T & 0 & 0 \\
 0 & -i \nu _{{dn}} \mathbf{I} & 0 & 0 \\
 -i \nu _{{A}} & 0 & -i \zsA \nuA & c_{{sn}} \mathbf{k}^T \\
 -\frac{\epsilon Z c_{{IA}}^2}{c_{{sn}}} \mathbf{k} & 0 & \frac{ \left(c_{{IA}}^2+c_{{si}}^2\right)}{c_{{sn}}} \mathbf{k} & -i \left(\nuA+\nuin\right)\mathbf{I} \\
\end{array}
\right)
\end{equation}
Just like before, the dust sector has a zero eigenvalue (the dust-acoustic wave has vanishing velocity at lowest order). We would like to determine whether the interaction terms modify the ion-acoustic dispersion relation sufficiently for it to have a zero eigenvalue at some wavelength. This is given by solutions to the equation:
\begin{equation}
    (\cia^2+\csi^2)\mathbf{k}^2=-\zsA\nuA(\nuA+\nuin)
\end{equation}
This equation does not have real roots unless $\zsA<0$, which would imply that the stopping time of the dust due to drag by ion adsorption \textit{decreases} as ion density increases. We therefore do not expect to see a resonant instability. (It is worth noting that if we instead considered the case of elastic dust-ion collisions, the resulting equation has a single root at $|\mathbf{k}|=0$.)

Computing growth rates of modes for the exact dust-ion system numerically using the parameters given in Section \ref{sec:ppdparams}, we find no mode with positive growth rate to within numerical error.

\section{Applications: Laboratory dusty plasmas}
\label{sec:lab}
The parallels drawn above raise the question of whether other resonant instabilities, namely the acoustic RDI and the two-stream instability, might appear under conditions similar to those of the filamentary instability if a nonzero dust velocity is added. In addition, as described in Section \ref{sec:astro}, the acoustic RDI and DIA two-stream instability are expected to play a role in diverse and often overlapping ranges of astrophysical environments. As such, laboratory experiments probing their behavior and evolution, either as separate or co-evolving instabilities, would inform the modelling of various astrophysical phenomena. Below we investigate this possibility by extending the model of Section \ref{sec:filament}, and by considering means of producing relevant conditions in a laboratory environment.

\subsection{Extending the model of Wang et. al.}
We consider a model extending that of Wang et. al. to include neutral dynamics. As such, we take the same assumptions as above but do not omit the neutral continuity and momentum equations. We take the same physical parameters as before. For the sake of consistency with the above RDI and two-stream models and with astrophysical models described below, we assume the dust to be accelerated relative to the neutrals by a uniform force, with both the dust and ions reaching equilibrium streaming velocities by balancing of drag forces.

\begin{figure*}
\subfloat[]{
\includegraphics[width=0.41\textwidth]{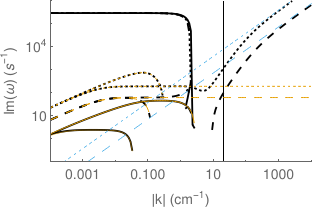}
}
\subfloat[]{
\includegraphics[width=0.41\textwidth]{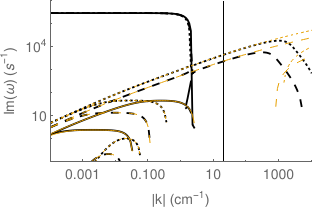}
}
\subfloat{
\includegraphics[width=0.18\textwidth]{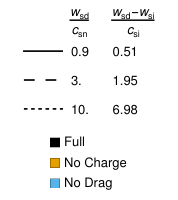}
}
\caption{Growth rates for the laboratory model, the idealized two-stream instability, and the model without ions or dust charge to demonstrate the acoustic RDI with wavevector (a) parallel to streaming (b) at the resonant angle for the acoustic RDI. The vertical line indicates $|k|=1/\lambda_{in}$.}
\label{fig:lab}
\end{figure*}

\begin{figure*}
    \includegraphics[width=0.75\textwidth]{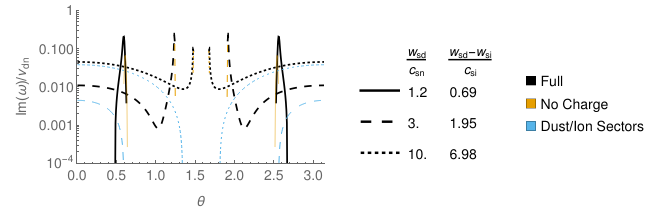}
    \caption{Growth rate of the fastest growing mode in the laboratory model unmodified, with the neutral matrix blocks removed to highlight the two-stream mode, and with the ion matrix blocks and dust charge removed to highlight the acoustic RDI at $|k|=1/\lambda_{in}$, with varied angle and streaming velocity. We see that the fastest growing mode may be either similar to the acoustic RDI or two-stream instability depending on angle and streaming velocity.}
    \label{fig:lab_angle}
\end{figure*}

The numerically calculated linear growth rates for this system are plotted in Figure \ref{fig:lab}, alongside those from Section \ref{sec:2S} and those for a neutral system of dust and gas with the same parameters. These growth rates can only be expected to be valid for wavelengths much longer than the dust spacing ($\nd_0^{-1/3}=0.0046$cm), the Debye length, and the viscous scale. Both the neutrals and ions collide with neutrals dramatically more frequently than with ions or dust, such that $\lambda_{in}=0.049$cm (calculated from Wang et. al.'s $\nuin$) and $\lambda_{nn}=0.015$cm (from the kinetic diameter of argon\cite{bakerMembraneTechnologyApplications2012}) are the relevant viscous scales. A vertical line has been included in Figure \ref{fig:lab} to indicate this limit. Figure \ref{fig:lab_angle} demonstrates that the growth rate of the unstable modes resembles that of either the acoustic RDI or two-stream instability depending on propagation angle.\\

\subsubsection{Acoustic resonant drag instability}
If we assume the system to be neutral, taking only the neutral and dust momentum and continuity equations, the equations become equivalent to those considered by Hopkins and Squire (hereafter HS) in their paper on the acoustic RDI\cite{hopkinsResonantDragInstability2018}. In those circumstances, one expects to find HS's acoustic RDI modes\cite{hopkinsResonantDragInstability2018}, and for such modes to persist with the addition of ions and dust charge. As 
 can be seen in Figure \ref{fig:lab}, modes qualitatively resembling those discussed by HS are visible, and the growth rates for the ionized and non-ionized cases match at long wavelengths.

\subsubsection{Two-stream instability}
If neutrals and drag are ignored, the system reduces to equation \ref{equ:2s}, which supports a dust-ion-acoustic two-stream instability. As a result, a similar two-stream mode should appear in this model system, with growth rates modified by drag and interaction with neutrals. This can be seen in Figures \ref{fig:lab} and \ref{fig:lab_angle}, in which modes resembling the two-stream mode appear at short wavelengths.

\subsubsection{Experimental accessibility}
\begin{figure*}
\includegraphics[width=\textwidth]{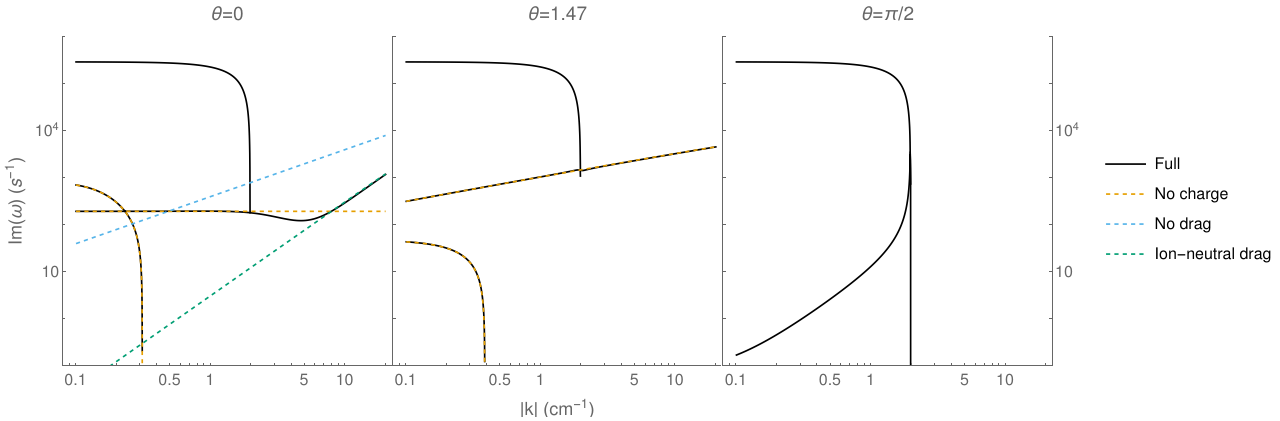}
\caption{Growth rates for the laboratory model, the model without drag, and the model without ions or dust charge, with wavevector (left to right) parallel to streaming, at the resonant angle for the acoustic RDI, and perpendicular to streaming. $w_s=2.45\times10^5$cm/s$=10\csn=7.0\csi=0.72\sqrt{\cia^2+\csi^2}$ and $|k|$ is between $0.1$cm$^{-1}$ and $1/\lambda_{in}=20.4$cm$^{-1}$. A model including only drag on the ions by a static neutral background is included to demonstrate a shift in the two-stream growth rate due to this effect.}
\label{fig:experiment}
\end{figure*}

Growth rates for the unstable modes in this laboratory model are shown in Figure \ref{fig:experiment} for wavelengths between the viscous scale $\lambda_{in}=0.049$cm and $10$cm for a dust-neutral streaming velocity of $7.0\csi$, and a corresponding equilibrium ion-neutral streaming velocity of $0.02\csi$. It can be seen that the filamentary mode, two-stream instability, and RDI are all accessible under these conditions, and might each be isolated by choice of experimental geometry to limit wavelength or propagation angle.

\subsection{Sheath levitation}
A system where all three of these modes may plausibly be probed is dust levitated in the sheath near a negatively charged surface. In such a system, the dust is held at an equilibrium distance from the boundary by a balance of electrostatic repulsion, gravity, and ion drag. Under such conditions, the dust may be made to levitate in regions where the ion flow velocity can be an order of magnitude larger than the ion thermal velocity and several times the ion-acoustic velocity.\cite{vladimirovEquilibriumLevitationDust2000} While the planar geometry of a dust cloud levitated in such a fashion may limit the viable directions of propagation of unstable modes, this is mitigated by the high streaming velocity extending the angle range of unstable modes to close to perpendicular to the ion flow (and nearly tangent to the plane). The maximum propagation angle of the DIA two-stream instability is $\arccos(c_{si}/\wsi)$, which is also the resonant angle for the dust/ion acoustic RDI. The filamentary mode might also be excited, normal to the ion flow. As the growth rates of these modes peak at different angles and wavelengths, each mode might be isolated by either limiting the wavelength (by constraining discharge geometry) or by limiting propagation angle (by tuning dust cloud thickness through dust grain size distribution or similar). A possible source of difficulty is the modification or suppression of these modes when the growth rate is slower than the ion crossing time $\tau=h/u_i$, where $h$ is the cloud width. Such behavior is not captured in the above homogeneous model.

\subsection{Dust acceleration}
Another possible means of producing these streaming instabilities in the laboratory is accelerating particles electrostatically through a static medium. This presents particular challenges for the acoustic RDI, as the dust must experience a restoring force maintaining its velocity over several stopping distances. For example, under the parameters considered in the above laboratory model, the electric field gradient required to maintain the dust traveling at the neutral sound speed is $0.82$kV/cm (substantially higher than the breakdown voltage), which would need to be present over distances much longer than $\csn/\nudn=4.0$cm. Producing such conditions in a plasma environment would present challenges due to sheath effects and resultant nonuniform potential profiles, such that is may not be possible to simultaneously probe both plasma and neutral streaming instabilities with this approach.

If only the acoustic RDI is considered, meaning the background gas may remain in a neutral state, producing a uniform electric field becomes simple, and electrostatic acceleration becomes more viable. The electric field required to balance (Epstein) drag is
\begin{equation}
    E=\sqrt{\frac{8}{\pi\gamma}(1+a_\gamma)}\frac{\rho \md c_s w_s}{eZ\brhod R_d}
\end{equation}
As long as this is kept below the breakdown voltage of the gas, the remaining constraint is a sufficiently long chamber for the time of flight to be much longer than the stopping time. Without a plasma to charge the dust, some other method of charging must be used. For solid dust grains, possible methods include contact with an electrode and UV irradiation.\cite{sternovskyChargingDustParticles2001} Another possibility is the production of charged droplets using an electrospray device.\cite{princeIonicLiquidsElectrospray2012,rosell-llompartGenerationMonodisperseDroplets1994}

\section{Applications: Astrophysical dusty plasmas}
\label{sec:astro}
With the above analysis placing two-stream instabilities and acoustic RDIs within the same framework, and having demonstrated that both modes are applicable within the same laboratory system, we consider their appearance and relative growth rates in several astrophysical dusty plasma environments.

\subsection{Comparing growth rates}
\label{sec:comparison}
The environments proposed to support acoustic RDIs are often at least partially ionized\cite{hopkinsResonantDragInstability2018}, and dust generically picks up a charge in a range of astrophysical environments. Therefore environments in which acoustic RDIs occur will in general plausibly support dust-ion two-stream instabilities. As a means of determining a heuristic for which of these effects is dominant, we derive a sufficient condition for the two-stream instability to have a comparable or faster growth rate.

We focus on the dominant acoustic RDI mode, the "quasi-drift" mode, whose growth rate at resonance is provided by HS.\cite{hopkinsResonantDragInstability2018} We use their result for short wavelength ($c_s t_s k_{||}\gg \frac{\mu+1}\mu$) as this matches conditions in the following subsections, and out of convenience. Assuming $\zwn$ and $\zsn$ are order unity, and that $\hwsd\equiv\frac\wsd{c_s}\gtrsim1$, the growth rate is bounded by the inequality:
\begin{equation}
    \Im(\omega)\gtrsim \left(\frac{\mu c_s t_s k}{16}\right)^{1/3}t_s^{-1}\equiv\gamma_{QD}
\end{equation}
As per equation \ref{equ:2Sgrowth}, the growth rate at resonance of the two-stream instability is (ignoring drag and assuming $\csi\lesssim\cia$)
\begin{equation}
    \Im(\omega)\approx \frac{\sqrt{3}(\cda\cia)^{2/3}k}{2^{4/3}(\cia^2+\csi^2)^{1/6}}\sim\frac\mi\md\epsilon Z^2 \cia k\equiv\gamma_{2S}
\end{equation}

It is important to note that this is only valid when the growth rate is substantially faster than the collision rates experienced by both dust and ions.
Comparing these and assuming Epstein drag, we find that the two-stream instability is faster than the quasidrift mode for
\begin{equation}
\begin{split}
    |\alpha-1|\sqrt\alpha|Z|\gtrsim
    &\left[\frac{8\pi\sqrt\gamman}9\sqrt\frac\mi\mn\left(\frac{T_n}{T_e}\right)^{3/2}\right]\times\\
    &(1+a_\gamma\hwsd^2)\frac{\Rd^4\nd n_g}{k^2}
\end{split}
\end{equation}
where $n_g$ is the density of the (possibly ionized) gas in the RDI. The product in brackets is generally near order unity, so the condition for the two-steam instability to have maximum growth rate at least comparable to that of the acoustic RDI is
\begin{equation}
    |\alpha-1|\sqrt\alpha |Z|\gtrsim\frac{\Rd^4 \nd n_g}{k^2}(1+a_\gamma\hwsd^2)
    \label{equ:inequality}
\end{equation}

We will see below that this provides a reasonably effective approximation of the "crossover" wavelength below which the two-stream instability begins to overtake the acoustic RDI.

\subsection{Weakly ionized environments}
Work has been done analyzing the effect of the acoustic RDI\cite{hopkinsResonantDragInstability2018} and the DIA two-stream instability\cite{havnesMotionDestructionGrains1980,shanStreamingInstabilitiesMulticomponent2008,rosenbergModifiedTwostreamInstabilities1995} on dust dynamics in weakly ionized media. Here we consider the overlap of these thus far separately studied phenomena.

\subsubsection{Protoplanetary disks}
 Due to the particular interest dust instabilities hold for planet formation in proplanetary disks, and the ready comparison of this environment to the weakly ionized laboratory plasma discussed above, we chose to focus on such an environment for numerical calculation of growth rates. We find modes corresponding to acoustic RDIs and two-stream instabilities. The physical relevance of these instabilities in protoplanetary disks is limited by their relatively low dust streaming velocities.\cite{hopkinsResonantDragInstability2018} Here we take the same model as Section \ref{sec:astroion}.
\paragraph{System parameters}
\label{sec:ppdparams}

We base our system parameters off of the protoplanetary disk model given  Okuzumi's 2009 paper on grain charging in protoplanetary disks.\cite{okuzumiELECTRICCHARGINGDUST2009} This paper considers a minimum-mass solar nebula, and calculates the dust charge and ionization ratio that would occur for varied dust grain size, assuming the dust particles to be fractal aggregates. (We assume $N=100$ monomers per particle.) Working within this model, we consider a region within the disk with $r=5AU$ and $z/H=1$, where $H(r)$ is the gas scale height, which fixes temperature, and neutral and dust number densities. We calculate the dust charge and ionization fraction assuming Okuzumi's formulae and value for the ionization rate at this location. We then adjust the ionization rate used in our model to match these values for varied streaming velocity. The parameters in our model not taken from that of Okuzumi are $\gamma=5/3$ and $\alpha_N=5.315\cdot a_0^3$ (corresponding to molecular hydrogen\cite{drainePhysicsInterstellarIntergalactic2011}).

\begin{figure*}
\subfloat[]{
\includegraphics[height=0.3\textwidth]{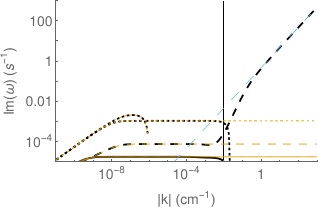}
}
\subfloat[]{
\includegraphics[height=0.3\textwidth]{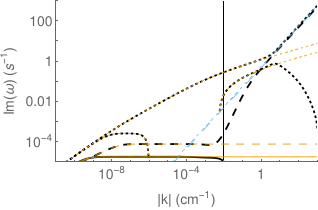}
}\\
\subfloat[]{
\includegraphics[height=0.3\textwidth]{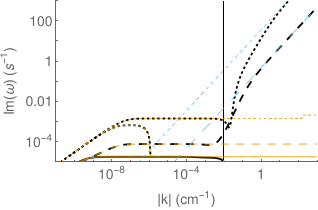}
}
\subfloat{
\includegraphics[height=0.3\textwidth]{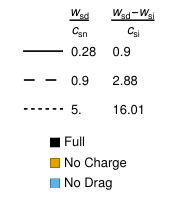}
}
\caption{Growth rates for the protoplanetary disk model, the idealized two-stream instability, and the model without ions or dust charge to demonstrate the acoustic RDI with wavevector (a) parallel to streaming (b) at the resonant angle for the acoustic RDI (c) at the resonant angle for the two-stream instability. The vertical line indicates $|k|=1/\lambda_{nn}$.}
\label{fig:ppd}
\end{figure*}

 As in the laboratory case, both ions and neutrals primarily collide with neutrals, such that the viscous scale is approximately $\lambda_{nn}=105cm$ or $\lambda_{in}=75cm$. The dust spacing is $\nd^{-1/3}=1.61cm$. We therefore take the minimum scale at which this model is valid to be set by viscosity to approximately $\lambda_{nn}$, which is indicated with a vertical line in Figure \ref{fig:ppd}.

\begin{figure}
    \includegraphics[width=0.5\textwidth]{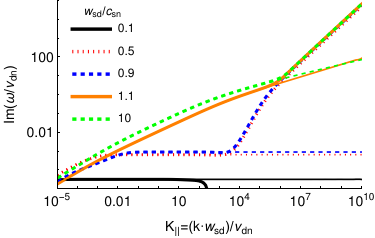}
    \caption{Growth rate of fastest growing mode at the acoustic RDI resonant angle for Hopkins and Squire's non-ionized acoustic mode (thin lines, reproducing the thick lines in their figure 3 top right)\cite{hopkinsResonantDragInstability2018} and with ions and dust charge included (thick lines). In this case, $\mu=0.01$ as per their figure 3, with all other parameters taken from our disk model.}
    \label{fig:ppd_HS}
\end{figure}

\paragraph{Acoustic RDIs}
As can be seen in Figure \ref{fig:ppd_HS}, we successfully replicate HS's acoustic RDIs for the Epstein drag case. However, we see an additional (two-stream) instability that appears at short wavelength, as discussed below. The growth rates of all unstable modes are shown at varied angle and streaming velocity in Figure \ref{fig:ppd}. Several of the visible modes are well-approximated by the equations with ions and dust charge removed, suggesting these to be the same acoustic RDIs. At long wavelength, these modes appear unaffected by charged species, while they are suppressed by them at short wavelength, where the two-stream instability becomes relevant. This effect is below the viscous cutoff in our model, but may be relevant in other environments.

\paragraph{Two-stream instability}
When the drag and ionization terms are dropped, the equations governing this system become those considered in Section \ref{sec:2S}. As can be seen in fig \ref{fig:ppd}, the growth rate of the idealized two-stream instability qualitatively matches that of the observed instability, and the instability appears when the streaming velocity exceeds the ion sound speed as expected. The two-stream instability is only dominant at wavelengths near or smaller than the viscous scale in this model. However, this successfully demonstrates the possibility of two-stream instabilities appearing in weakly ionized dusty plasmas in astrophysical settings. The criterion of equation \ref{equ:inequality} predicts a critical wavelength of $|k|\sim 10^{-3}$cm$^{-1}$, which is comparable to what is seen in Figure \ref{fig:ppd}.

\subsubsection{Shocked ISM}
\begin{figure*}
\subfloat[]{
\includegraphics[width=0.5\textwidth]{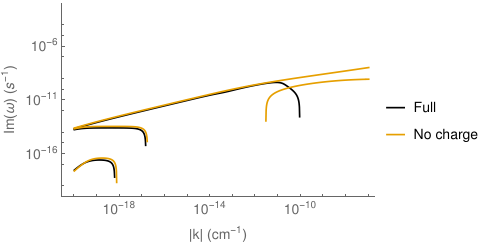}
}
\subfloat[]{
\includegraphics[width=0.5\textwidth]{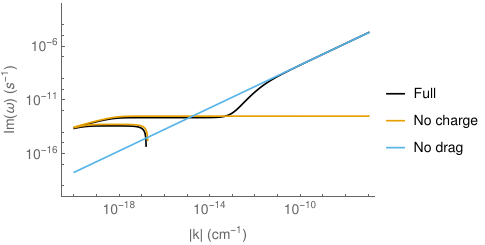}
}
    \caption{Calculated growth rates in shocked ISM of the two-stream instability and acoustic RDIs at the resonant angle for a) the acoustic RDI ($\theta=0$) and b) the two-stream instability ($\theta=0.73$) in shocked ISM. Growth rates with drag or dust charge turned off are included for comparison, isolating the two-stream instability and acoustic RDI. The two-stream instability is not present at $\theta=0$.}
    \label{fig:ism}
\end{figure*}

A shock moving through dusty ISM may create supersonic dust streaming velocities for timescales shorter than the dust stopping time. Havnes has analyzed the growth rate of the resulting two-stream instability using a kinetic model, finding it should strongly affect dynamics prior to the dust velocity equilibrating to the surrounding medium.\cite{havnesMotionDestructionGrains1980} Following formulas used by Havnes, we take $Z=2.51R_dk_BT/e^2$, $T=\frac\mn{3k_B}\wsd^2$, $\mn=\mn=1.3m_p$, $\mu=0.0025$, $\xI=0.001$, $R_d=10^{-6}$cm, $\brhod=3$g/cm$^3$, $\nn=100$cm$^{-3}$. We assume Epstein dust-neutral drag, Coulumb dust-ion drag, and use equation \ref{equ:sigmav} for ion-neutral drag, taking $\left(\frac{\alpha_N}{a_0^3}\right)=4.5$. We consider the case with $\wsd=50$km/s. The growth rates for instabilities in this system are shown in Figure \ref{fig:ism}. We see the expected two-stream instability at short wavelength, as well as an acoustic RDI at long wavelength, which is seen in our analysis but not in that of Havnes due to our inclusion of collision terms. The critical wavelength predicted by equation \ref{equ:inequality} is $|k|\approx2\times 10^{-14}$cm$^{-1}$, which is comparable to the crossover seen in Figure \ref{fig:ism}.

There are several caveats to these results. The first is that the mean free path is $\sim10^{12}$cm, such that in the regime in which the two-stream instability is found, it will be modified by viscous effects not included in our model. In addition, plugging the above parameters into the expression given by Havnes for the approximate growth rate, we find the result to be $\sim10^3$ times larger than ours. However, our goal is to demonstrate the appearance of the acoustic RDI under conditions in which the DIA two-stream instability is known to appear, so such factors concerning the two-stream instability do not substantively modify our conclusions. Concerning the acoustic RDI, the growth rate is also seen to be highest at viscous scales, which will modify the actual growth rates. In particular, the instability is of relevance when its growth rate is faster than the collision rate ($\nudn=4.3\times 10^{-10}$s$^{-1}$) i.e. before the dust is slowed down, which can be seen to occur essentially at viscous scales.

\subsection{Strongly ionized environments}
Intuitively, one would expect the two-stream instability to become more relevant in strongly ionized environments, where electrostatic effects are more relevant. Here we consider two such systems suggested by HS to support acoustic RDIs, dusty winds found in the vicinity of active galactic nuclei (AGNs) and planetary nebulae (PNes).\cite{hopkinsResonantDragInstability2018} In both cases, magnetic fields are ignored. Our model is substantially simpler than those used in the preceding sections, as the gas may be treated as a single ionized species.

We choose to ignore adsorption of ions onto dust, treating ion-dust collisions as Epstein drag. We find the viscous scale (at the boundary of validity of our model and maximum growth rate of the instabilities) to be set by the ion-ion mean free path, approximated by HS as $\lambda_{mfp}\sim(T/10^4\mathrm{K})^2(n_g/\mathrm{cm}^{-3})^{-1}10^{12}\mathrm{cm}$.\cite{hopkinsResonantDragInstability2018} In both cases, the fastest drag term is that on the dust from the ions, which we see is substantially slower than either instability (assuming Epstein drag), making the calculations from Section \ref{sec:comparison} applicable. Conveniently, since there is only one characteristic wave speed, $\sqrt{\cia^2+\csi^2}$, the resonant angle is the same for both modes.

\subsubsection{AGN Winds}
\begin{figure*}
\subfloat[]{
\includegraphics[width=0.5\textwidth]{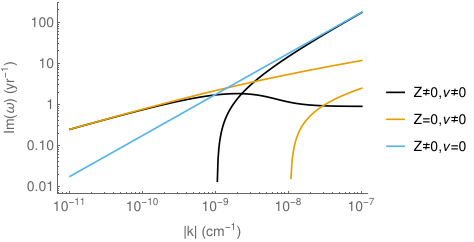}
}
\subfloat[]{
\includegraphics[width=0.5\textwidth]{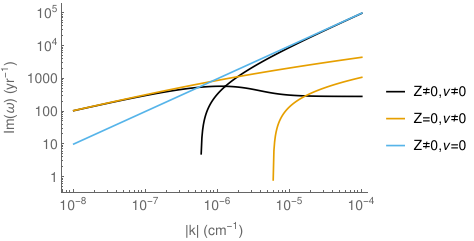}
}
\caption{Calculated growth rates of two-stream and acoustic RDIs at the resonant angle in AGN outflows. Growth rates with drag or dust charge turned off are included for comparison, isolating the two-stream instability and acoustic RDI. Parameters are approximated from Sarangi et. al.\cite{sarangiDustFormationAGN2019} and Tazaki et. al.\cite{tazakiDustDestructionCharging2020} for polar coordinates (a) $r\sim10^{19}$cm, $\theta\sim30^\circ$ and (b) $r\sim10^{17.5}$cm, $\theta\sim75^\circ$.}
    \label{fig:agn}
\end{figure*}

As mentioned by HS, the vicinity of an AGN contains supersonic dusty outflows.\cite{hopkinsResonantDragInstability2018} They expect these to support acoustic RDIs on a broad range of scales, giving the characteristic time of the instability as $10-100$ hours for wavelengths near a viscous scale of $10^7-10^8$cm. It is important to note that they state this to be in the "mid-$k$" regime, while we instead find that the minimum wavelength is firmly in the "high-$k$" regime ($c_s t_s k_{||}\gg \frac{\mu+1}\mu$), perhaps due to differences in assumed dust density. We estimate dust density, gas density, and gas temperature from simulation data in Sarangi et. al.\cite{sarangiDustFormationAGN2019} We consider two locations in their model, at $r\sim10^{19}$cm, $\theta\sim30^\circ$, and at $r\sim10^{17.5}$cm, $\theta\sim75^\circ$ in polar coordinates. In similar fashion to HS, we take $\frac{w_s}{c_s}\sim100\left(\frac{r}{0.3pc}\right)^{-1/2}$. We assume silicate dust ($\brhod=1.4$g/cm$^3$) of radius $0.01\mu$m. Calculations by Tazaki et. al. indicate that silicate dust grains of this size in AGN conditions are charged by photoelectric emission to $\sim5\times10^{1-2}$V, leading to possible Coulumb explosion of the grain above $\sim100$V.\cite{tazakiDustDestructionCharging2020} Taking the dust to be charged to $\sim100$V, the grain charge is $\sim+500e$, corresponding to $Z=-500$ in our (now somewhat inconvenient) notation. As shown in Figure \ref{fig:agn}, we find that the two-stream instability grows more rapidly than the acoustic RDI for wavelengths $\sim10^{0-2}\lambda_{mfp}$, evolving on timescales as short as minutes for wavelengths around hundreds of meters. Equation \ref{equ:inequality} predicts this to occur below wavelengths of $k\approx 4\times 10^{-10}$cm$^{-1}$ and $k\approx 2\times 10^{-7}$cm$^{-1}$ for the first and second locations respectively, well within an order of magnitude of the observed result.

\subsubsection{Planetary Nebulae}
\label{sec:PNe}
\begin{figure*}
\subfloat[]{
\includegraphics[width=0.5\textwidth]{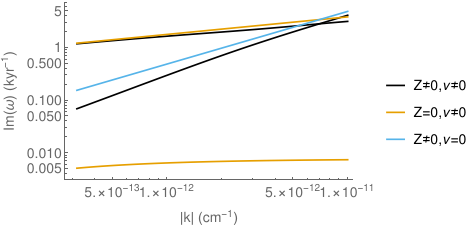}
}
\subfloat[]{
\includegraphics[width=0.5\textwidth]{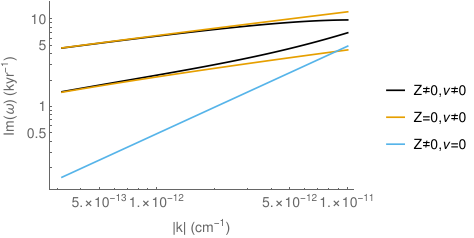}
}%
\caption{Calculated growth rates of two-stream and acoustic RDIs at the resonant angle in a planetary nebula. Growth rates with drag or dust charge turned off are included for comparison, isolating the two-stream instability and acoustic RDI. Parameters are approximated from Borkowski and Harrington\cite{borkowskiGrainheatedDustyPlanetary1991} for (a) $w_s=10$km/s and (b) $w_s=100$km/s.}
    \label{fig:pne}
\end{figure*}

HS briefly mention the possibility of acoustic RDIs appearing in the stellar winds in planetary nebulae.\cite{hopkinsResonantDragInstability2018} In order to calculate estimates from concrete parameters, we consider the planetary nebula IRAS 18333-2357 (GJJC1) as modeled in the paper by Borkowski and Harrington concerning grain-heating of the nebula.\cite{borkowskiGrainheatedDustyPlanetary1991} We use the parameters referenced as Model A within the paper. The nebula is described as being composed largely of singly ionized helium, with $n_g\sim8$cm$^{-3}$ and $T\sim10^4$K. The dust to gas mass ratio is given as $0.311$, the mean dust radius as $r=0.0548\mu$m, and grain density as $\brhod=2.26$g/cm$^{-3}$ (carbon), yielding a dust density of $\nd=10^{-8}$cm$^{-3}$. The grain potential due to photoelectric emission is calculated to vary with radial position but is of order $\sim5$V, which would correspond to a dust charge of $+190e$ ($Z=-190$). Harrington indicates that winds within the nebula are expected to be $\sim10^3$km/s. We consider dust streaming velocities of $10$km/s and $100$km/s, finding that near the viscous scale the two-stream instability grows on timescales of hundreds of years, while the acoustic RDI may have a growth rate comparable to or exceeding this depending on the streaming velocity, as shown in figure \ref{fig:pne}. At wavelengths longer than $~10\lambda_{mfp}$, the acoustic RDI is dominant, with the growth rates becoming comparable near the scale predicted by equation \ref{equ:inequality} ($|k|\approx 2\times 10^{-12}$cm$^{-1}$ and $|k|\approx 1\times 10^{-11}$cm$^{-1}$ for velocities of $10$km/s and $100$km/s respectively). As the lifetime of a planetary nebula is on the order of $10^4$ years, both instabilities may play a role in the evolution of such a system. It is possible that the two-stream instability is the dominant mode at scales below the mean free path, but a more detailed model would be required to study this.

\section{Conclusion}
The primary results of this paper are the extension of SH's perturbative framework for resonant drag instabilities to a broader class of dust-fluid instabilities, and the proposal of a consequent overlap in the expected regimes, both laboratory and astrophysical, in which they appear. Due to the general tendency for scale separation within dust-fluid systems, their linearization is likely to produce nearly block triangular matrices. The inclusion of dust streaming allows the spectrum to be tunable to produce a nearly defective eigenvalue, and the appearance of such resonant instabilities would be expected to be a ubiquitous phenomenon regardless of the form of interaction. We note that in the case of electrostatic interactions between dust and ions, the smallness of $\mi/\md$ allows the action of the ions on the dust to be treated as first order in the same way that the smallness of $\mu$ allows the action of the dust on ions or neutrals to be treated as first order in the case of collisional interaction.

We use this to apply a perturbative treatment to the filamentary ionization instability and to the dust-ion-acoustic two-stream instability. The dust-ion-acoustic two-stream instability and the filamentary instability are both considered in a multifluid model of a homogeneous unmagnetized dusty plasma, which should also be able to support acoustic RDI modes. As such, we consider the relative growth rates of these modes in various environments. In particular, we show that the acoustic RDI and DIA two-stream instability can often be expected to appear concurrently in astrophysical environments, particularly in strongly ionized regimes, and that all three modes should be accessible in laboratory, possibly within a single system. We therefore propose this as an area of interest for future laboratory astrophysics.

These results suggest several possible directions of further work. More robust fluid or kinetic models may be applied to probe these instabilities in magnetic fields, at viscous scales, or with the inclusion of a spectrum of dust sizes\cite{squireAcousticResonantDrag2022}, all of which would be factors relevant to the discussed astrophysical environments. Simulations of astrophysical dusty plasmas often rely on a (magneto)hydrodynamic description of the gas which, even when dust charge is included, does not account for the electrostatic interaction of the gas with the dust. It is possible that the inclusion of electrostatic effects may be necessary in simulating ionized environments due to the existence of two-stream instabilities. The nonlinear evolution of dust-ion two-stream instabilities in the aforementioned astrophysical systems and its contribution to turbulent flows in such environments provide a motivation for such simulations. Further, the possible simultaneous presence of the  modes discussed above within one system leads to the possibility of nonlinear interaction between them, presenting an open direction of numerical and experimental study of dusty plasma phenomena relevant to astrophysics.

\begin{acknowledgements}
We would like to thank Yevgeny Raitses and Nirbhav Chopra for productive discussions and useful suggestions concerning experimental techniques and parameters, and Eric Moseley and Bruce Draine for helpful input concerning dusty astrophysical environments.

This work was supported by the U.S. Department of Energy under contract number DE-AC02-09CH11466. The United States Government retains a non-exclusive, paid-up, irrevocable, world-wide license to publish or reproduce the published form of this manuscript, or allow others to do so, for United States Government purposes.
\end{acknowledgements}

\appendix

\section{List of symbols}
\label{sec:symbols}
The location in the paper where the symbol is defined is included where relevant. Some symbols that are used only immediately adjacent to their definitions are omitted from this table.

For brevity, symbols that may be associated with more than one species are denoted with subscript $j$ or $k$, which may be replaced with $d,\ i,\ n,\ e$ for dust, ion, neutral, or electron respectively. Subscripts in parentheses are optional, and subscripts that may take multiple values are enumerated with commas.

\begin{longtable}{l|l|l}
$ A$ & Dust sector of zeroth order matrix & Eq. \ref{equ:T0T1}\\
$ A_W$ & Parameter to match Wang et. al.\cite{wangIonizationInstabilitiesResonant2001} & Sec. \ref{sec:Wang-model}\\
$ C$ & Dust-fluid sector of zeroth order matrix & Eq. \ref{equ:T0T1}\\
$ F$ & Fluid sector of zeroth order matrix & Eq. \ref{equ:T0T1}\\
$ \mathcal{P}$ & Parity reversal operator & App. \ref{sec:PT}\\
$ R_d$ & Dust grain radius & Sec. \ref{sec:model}\\
$ \mathcal{T}$ & Time reversal operator & App. \ref{sec:PT}\\
$ T_{(0,1)}$ & (Zeroth, first order) matrix & Eq. \ref{equ:linear}\\
$ T_j$ & Temperature\\
$ Z$ & Dust charge ($Z>0$ for \textit{negative} charge) & Sec. \ref{sec:model}\\
$ a$ & Dust acceleration & Sec. \ref{sec:model}\\
$ a_0$ & $\hbar/m_e e^2$ & App. \ref{sec:nidrag}\\
$ a_{\gamma(j)}$ & $9\pi\gamma_{(j)}/128$ & App. \ref{sec:Epsteindrag}\\
$ \cda$ & Dust-ion-acoustic velocity & Eq. \ref{equ:speeds}\\
$ \cia$ & Ion-acoustic velocity & Eq. \ref{equ:speeds}\\
$ c_{s(j)}$ & Sound speed & Eq. \ref{equ:speeds}\\
$ e$ & Elementary charge\\
$ k$ & Wave vector\\
$ k_{||}$ & Component of wave vector $||$ streaming\\
$ m_j$ & Species mass\\
$ m_p$ & Proton mass\\
$ n_j$ & Number density\\
$ t_{dA}$ & Adsorption time & Sec. \ref{sec:collisions}\\
$ t_{jk}$ & Collision time & Sec. \ref{sec:collisions}\\
$ u_j$ & Flow velocity\\
$ \hwsd$ & $\wsd/c_s$\\
$ w_{s(j)}$ & Streaming velocity & Sec. \ref{sec:model}\\
$ \vec{x}$ & Linear perturbation vector & Sec. \ref{sec:RDI}\\
$ x_I$ & Ionization ratio & Eq. \ref{equ:dimless}\\
$ \alpha$ & $1/(1-\epsilon Z)$ & Eq. \ref{equ:dimless}\\
$ \alpha_N$ & Neutral polarizability & App. \ref{sec:nidrag}\\
$ \gamma_{(j)}$ & Polytropic index\\
$ \epsilon$ & $n_d/n_i$ & Eq. \ref{equ:dimless}\\
$ \zeta_{sA}$ & Adsorption density coefficient & Eq. \ref{equ:colcoeffs}\\
$ \zeta_{ni}$ & Neutral-ion drag density coefficient & Eq. \ref{equ:colcoeffs}\\
$ \zeta_{si}$ & Dust-ion drag density coefficient & Eq. \ref{equ:colcoeffs}\\
$ \zeta_{sn}$ & Dust-neutral drag density coefficient & Eq. \ref{equ:colcoeffs}\\
$ \zeta_{wA}$ & Adsorption velocity coefficient & Eq. \ref{equ:colcoeffs}\\
$ \zeta_{wi}$ & Dust-ion drag velocity coefficient & Eq. \ref{equ:colcoeffs}\\
$ \zeta_{wn}$ & Dust-neutral drag velocity coefficient & Eq. \ref{equ:colcoeffs}\\
$ \theta$ & Angle between $\wsd$ and $k$ & Eq. \ref{equ:dimless}\\
$ \lambda_{jk}$ & Mean free path\\
$ \mu$ & Dust-gas mass density ratio & Eq. \ref{equ:dimless}\\
$ \nu_A$ & Adsorption rate & Sec. \ref{sec:model}\\
$ \nu_I$ & Ionization rate & Sec. \ref{sec:model}\\
$ \nu_{jk}$ & Collision rate & Sec. \ref{sec:model}\\
$ \nu_L$ & Ion loss rate & Sec. \ref{sec:model}\\
$ \xi_{A,F}^{L,R}$ & Left, right eigenvectors of $A$, $F$ & Sec. \ref{sec:RDI}\\
$ \rho$ & Gas mass density\\
$ \brhod$ & Dust grain material density\\
$ \Phi$ & Electric potential\\
$ \omega$ & Frequency\\
\end{longtable}

\section{Parameter values}
\label{sec:values}
Listed in tables \ref{tab:weak} and \ref{tab:strong} are various parameter values at equilibrium for the systems considered above.

\setcounter{table}{0}
\begin{table*}
\begin{tabular}{|c|c|ccc|c|}
\hline
 & Lab & PPD & & & ISM \\
 & & $\wsd=0.9\csi$ & $\wsd=0.9\csn$ & $\wsd=5\csn$ & \\
\hline
 $\wsd$ (cm/s) & $0$ & ${2.41\times10^{4}}$ & ${7.72\times10^{4}}$ & ${4.29\times10^{5}}$ & ${5\times10^{6}}$ \\
 $\wsi$ (cm/s) & $0$ & $-0.0666$ & $-0.240$ & $-3.73$ & $0$ \\
 \hline
 $m_d$ ($m_p$) & ${8\times10^{8}}$ & ${3.51\times10^{11}}$ & $''$ & $''$ & ${7.51\times10^{6}}$ \\
 $m_i$ ($m_p$) & $40$ & $24$ & $''$ & $''$ & $1$ \\
 $m_n$ ($m_p$) & $40$ & $2.34$ & $''$ & $''$ & $1.3$ \\
 \hline
 $n_d$ (cm$^{-3}$) & ${1\times10^{7}}$ & $0.240$ & $''$ & $''$ & ${4.33\times10^{-8}}$ \\
 $n_i$ (cm$^{-3}$) & ${2\times10^{9}}$ & $0.0772$ & $''$ & $''$ & $0.1$ \\
 $n_n$ (cm$^{-3}$) & ${1.33\times10^{16}}$ & ${2.56\times10^{12}}$ & $''$ & $''$ & $100$ \\
 \hline
 $T_n$ (eV) & $0.0250$ & $0.0108$ & $''$ & $''$ & $11.3$ \\
 $T_i$ (eV) & $0.0512$ & $0.0108$ & $''$ & $''$ & $11.3$ \\
 $T_e$ (eV) & $3$ & $0.0108$ & $''$ & $''$ & $11.3$ \\
 $\gamma$ & $1$ & ${5/3}$ & $''$ & $''$ & ${5/3}$ \\
 \hline
 $\nudn$ (s$^{-1}$) & ${5.60\times10^{3}}$ & $0.0771$ & $0.0866$ & $0.243$ & ${4.31\times10^{-10}}$ \\
 $\nudi$ (s$^{-1}$) & $3.47$ & ${}$ & ${}$ & ${}$ & ${1.13\times10^{-11}}$ \\
 $\nuni$ (s$^{-1}$) & $0.150$ & ${1.21\times10^{-10}}$ & $''$ & $''$ & ${1.33\times10^{-10}}$ \\
 $\nuL$ (s$^{-1}$) & ${5\times10^{5}}$ & ${}$ & ${}$ & ${}$ & ${}$ \\
 $\nuI$ (s$^{-1}$) & ${5\times10^{5}}$ & ${3.79\times10^{-4}}$ & ${6.69\times10^{-4}}$ & ${3.25\times10^{-3}}$ & ${}$ \\
 $\nuA$ (s$^{-1}$) & ${}$ & ${3.79\times10^{-4}}$ & ${6.69\times10^{-4}}$ & ${3.25\times10^{-3}}$ & ${}$ \\
 \hline
 $\brhod$ (g/cm$^3$) & $2.56$ & $0.14$ & $''$ & $''$ & $3$ \\
 $Z$ & $75$ & $0.317$ & $''$ & $''$ & $197.$ \\
 $r$ ($\mu$m) & $0.05$ & $1$ & $''$ & $''$ & $0.01$ \\
 \hline
 d/n drag & \text{Epstein} & \text{Epstein} & $''$  & $''$  & \text{Epstein} \\
 d/i drag & \text{Coulumb} & \text{Epstein} & $''$  & $''$  & \text{Coulumb} \\
 \hline
\end{tabular}

\caption{Values for various parameters at equilibrium for the weakly ionized systems considered above: laboratory dusty plasmas, protoplanetary disks, and shocked interstellar medium. Parameters unused in a particular model are left blank.}
\label{tab:weak}
\end{table*}

\begin{table*}
\begin{tabular}{|c|cc|cc|c|cc|cc|}
\hline
 &  AGN & & PNe & \\
 & $r=10^{19}$cm & $r=10^{17.5}$cm & $\wsd=10$km/s & $\wsd=100$km/s \\
 & $\theta=30^\circ$ & $\theta=75^\circ$ & & \\
\hline
 $\wsd$ (cm/s) & ${2.01\times10^{7}}$ & ${6.35\times10^{7}}$ & ${1\times10^{6}}$ & ${1\times10^{7}}$ \\
 \hline
 $m_d$ ($m_p$) & ${3.51\times10^{6}}$ & $''$ & ${9.31\times10^{8}}$ & $''$ \\
 $m_i$ ($m_p$) & $1$ & $''$ & $3.97$ & $''$ \\
 \hline
 $n_d$ (cm$^{-3}$) & ${1\times10^{-6}}$ & ${1\times10^{-4}}$ & ${1.06\times10^{-8}}$ & $''$ \\
 $n_i$ (cm$^{-3}$) & ${1\times10^{4}}$ & ${1\times10^{6}}$ & $8$ & $''$ \\
 \hline
 $T$ (eV) & $0.273$ & $0.0862$ & $0.862$ & $''$ \\
 $\gamma$ & ${5/3}$ & $''$ & ${5/3}$ & $''$ \\
 \hline
 $\nudi$ (s$^{-1}$) & ${1.80\times10^{-7}}$ & ${5.69\times10^{-5}}$ & ${4.49\times10^{-12}}$ & ${3.24\times10^{-11}}$ \\
 \hline
 $\brhod$ (g/cm$^3$) & $1.4$ & $''$ & $2.26$ & $''$ \\
 $Z$ & $-500$ & $''$ & $-190.$ & $''$ \\
 $r$ ($\mu$m) & $0.01$ & $''$ & $0.0548$ & $''$ \\
 \hline
 d/i drag & \text{Epstein} & $''$  & \text{Epstein} & $''$  \\
 \hline
\end{tabular}

\caption{Values for various parameters at equilibrium for the strongly ionized systems considered above: active galactic nucleus outflows and planetary nebula winds.}
\label{tab:strong}
\end{table*}

\section{PT-symmetry}
\label{sec:PT}
P- and T-symmetry, commonly discussed in the context of quantum mechanics, refer respectively to parity reversal and time reversal symmetries. Parity reversal, with operator $\mathcal{P}$, generally refers to the reversal of spatial coordinates and corresponding derivatives ($x\rightarrow -x$, $v\rightarrow -v$, $\nabla\rightarrow -\nabla$). Time reversal, with operator $\mathcal{T}$, refers to a similar operation on the time coordinate ($t\rightarrow-t$, $v\rightarrow-v$, $\partial_t\rightarrow-\partial_t$). In the context of matrix representations of linearized systems, $\mathcal{T}$ is simply the complex conjugation operator, and $\mathcal{P}$ is some matrix such that $\mathcal{P}^2=\mathbf{I}$.

A linearized system $i\dot{\vec x}=M\vec x$ is said to possess PT-symmetry if its matrix $M$ (avoiding the symbol $T$ used in other sections to prevent confusion) commutes with $\mathcal{P}\mathcal{T}$ i.e. $\mathcal{P}\mathcal{T}M=M\mathcal{P}\mathcal{T}$. Such matrices have the property that their eigenvalues are symmetric with respect to the real axis, meaning all of their eigenvalues are either real or in complex conjugate pairs.\cite{benderMakingSenseNonHermitian2007}

\subsection{Spontaneous symmetry breaking}
\begin{figure}
    \subfloat[Allowed]{
    \includegraphics[width=0.25\textwidth]{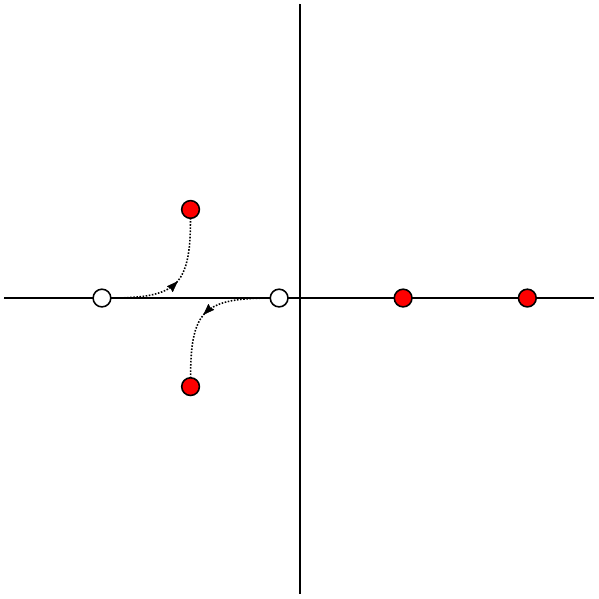}
    }
    \subfloat[Forbidden]{
    \includegraphics[width=0.25\textwidth]{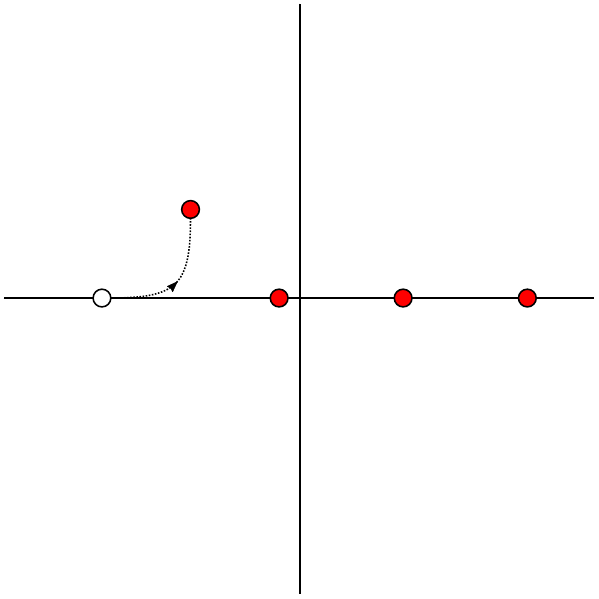}
    }
    
    \caption{In PT-symmetric systems, a real eigenvalue may only become complex by colliding with another real eigenvalue, with which it may move off the real line as a complex conjugate pair.}
    \label{fig:Krein}
\end{figure}

A symmetry is said to be spontaneously broken when, while $M$ respects the symmetry, its eigenvectors are not eigenvectors of the symmetry operator. In the case of PT-symmetry, this occurs when there exists a complex conjugate pair of eigenvalues, whose eigenvectors are then exchanged by the symmetry operator. This is precisely when the system exhibits instability ($Im(\omega)>0$), as has been noted for a number of fluid and plasma systems.\cite{qinKelvinHelmholtzInstabilityResult2019,fuPhysicsSpontaneousParitytime2020,qinKelvinHelmholtzInstabilityResult2019,qinSpontaneousExplicitParitytimesymmetry2021,zhangStructureTwostreamInstability2016} As illustrated in Figure \ref{fig:Krein}, the system can only move from a stable to an unstable regime of parameter space via the collision of two real eigenvalues which subsequently leave the real axis, a resonance known as a Krein collision.

\subsection{Two-stream instability}
The system described in Section \ref{sec:2S} possesses T-symmetry, as is manifest in all elements of $M$ being real. As such, it possesses a degenerate form of PT-symmetry, in which $\mathcal{P}=\mathbf{I}$ is a trivial transformation; the resulting formal properties nonetheless apply. $\mathcal{P}$ may be nontrivial for a different choice of field variables, such as density and the time derivative of density\cite{zhangStructureTwostreamInstability2016}, instead of density and velocity as used in this paper. As seen in Figure \ref{fig:PT-2S}, tracing eigenvalues over varied streaming velocity (and with $\mathbf{w_s}||\mathbf{k}$) shows two Krein collisions as expected: two real modes move together until a critical streaming value, beyond which they move into the complex plane before colliding and separating along the real axis at another critical velocity.

\subsection{Resonant drag instabilities}
\begin{figure*}
    \includegraphics[width=0.7\textwidth]{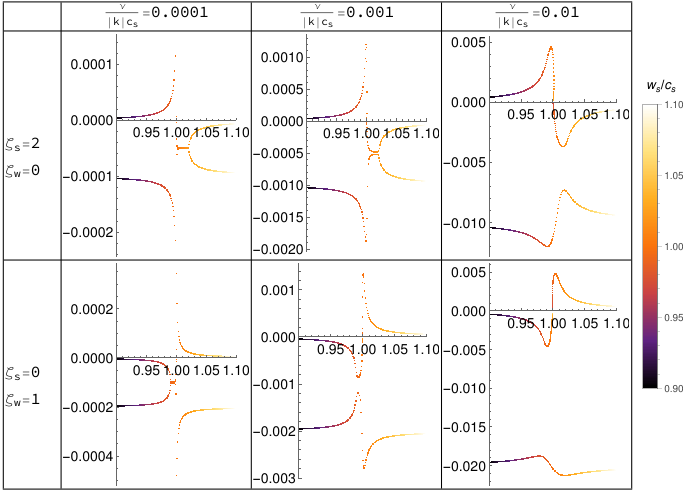}\\
    \caption{Plots of eigenvalues for the acoustic RDI in arbitrary units for varied streaming velocity with differing collision physics and drag. $\mu$ is fixed at $0.01$, and $\mathbf{k}||\mathbf{w_s}$.}
    \label{fig:PT-RDI}
\end{figure*}

\begin{figure}
    \subfloat[]{
    \includegraphics[width=0.5\textwidth]{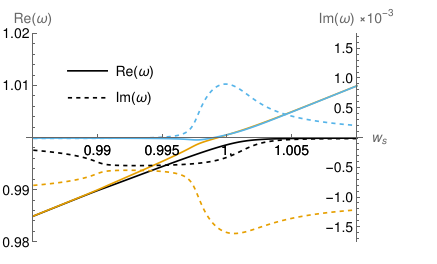}
    }
    
    \subfloat[]{
    \includegraphics[width=0.5\textwidth]{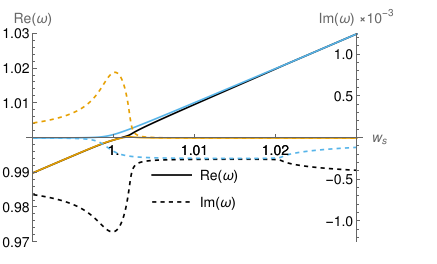}
    }%
    \caption{Real and imaginary components of eigenvalues for the acoustic RDI with varied streaming velocity and parallel propagation. Two choices of $\zeta_s,\ \zeta_w$ are shown: a) $\zeta_s=0,\ \zeta_w=1$ b) $\zeta_s=2,\ \zeta_w=0$. The drag coefficient $\nu$ was tuned to make the structural features easily visible. $\mu=0.01,\ \nu=0.0005$ in both cases, and units are normalized such that $k=c_s=1$.
    In both cases, the two bifurcations visible in Figure \ref{fig:PT-RDI} are present: a Krein-like resonance, where the real part of two eigenvalues are able to diverge once the imaginary components approach a constant, and a more complex interaction, where the onset of instability is associated with a near-resonance between a different pair of eigenvalues.}
    \label{fig:PT-RDI-split}
\end{figure}

At first glance, spontaneous PT-symmetry breaking seems inherently irrelevant to drag-induced instabilities such as RDIs, as the non-conservative nature of such systems breaks PT-symmetry explicitly \cite{qinSpontaneousExplicitParitytimesymmetry2021}. However, the parallels thus far demonstrated between RDIs and two-stream instabilities motivate consideration of the limit of weak drag, in which PT-symmetry is nearly unbroken and the dust sector possesses a nearly degenerate eigenvalue. Similarly to the plots of the two-stream instability in Figure \ref{fig:PT-2S}, we show in Figure \ref{fig:PT-RDI} eigenvalues for the acoustic RDI system with varied streaming velocity. While they are no longer symmetric with respect to the real axis, they nonetheless seem to possess a symmetry about a horizontal line below the real axis, and show two bifurcations of the eigenvalue, one of which is near the boundary of stability. Plots illustrating the details of the resonance and bifurcations are shown in Figure \ref{fig:PT-RDI-split}. A more detailed study of these phenomena, including the possibility of a perturbative approach to explain these similarities to conservative systems, is left for future work.

\section{Collision physics}
\label{sec:collision-physics}
Here we enumerate the collision times $t$ and linearization coefficients $\zeta$ (as defined in Section \ref{sec:collisions}) relevant to the models discussed in this paper.

\subsection{Epstein drag}
\label{sec:Epsteindrag}
For dust-neutral drag and for dust-ion drag when the ion kinetic energy is much larger than the grain surface potential, we assume an Epstein drag model. We also apply this model for drag due to ion adsorption onto dust.\cite{bainesResistanceMotionSmall1965,drainePhysicsDustGrains1979,hopkinsResonantDragInstability2018}
\begin{equation}
\begin{split}
\tdn=&\htdn\left(\frac{\nn_0}{\nn}\right)^{\frac{\gamman+1}{2}}\left(1+\agn\frac{|\ud-\un|^2}{\csn^2}\right)^{-\frac12}\\
\zeta_{sn}=&\frac{1+\gamma_n+2a_{\gamma n}\hat w_n^2}{2(1+a_{\gamma n}\hat w_n^2)}\\
\zeta_{wn}=&\frac{a_{\gamma n}\hat w_n^2}{1+a_{\gamma n}\hat w_n^2}\\
\end{split}
\end{equation}
\begin{equation}
\begin{split}
t_{d(A/i)}=&\htdi\left(\frac{\ni_0}{\ni}\right)^{\frac{\gammai+1}{2}}\left(1+\agi\frac{|\ud-\ui|^2}{\csi^2}\right)^{-\frac12}\\
\zeta_{s(A/i)}=&\frac{1+\gamma_i+2a_{\gamma i}\hat w_i^2}{2(1+a_{\gamma i}\hat w_i^2)}\\
\zeta_{w(A/i)}=&\frac{a_{\gamma i}\hat w_i^2}{1+a_{\gamma i}\hat w_i^2}\\
\end{split}
\end{equation}
We have used the following definitions (where $j=n,i$):
\begin{equation}
\begin{split}
\hat t_{dj}=&\left(\frac{\pi\gamma_j}{8}\right)^{\frac12}\frac{\brhod\Rd}{m_j n_{j0}{c_{sj}}_0}\\
c_{sj}^2=&\left(\frac{n_j}{{n_j}_0}\right)^{\gamma-1}{c^2_{sj}}_0\\
a_{\gamma j}=&\frac{9\pi\gamma_j}{128}\\
\hat w_j=&\frac{u_{d0}-u_{j0}}{c_{sj0}}\\
\end{split}
\end{equation}
$\gamma_j$ is the polytropic index of species $j$. We will always choose to have dust-ion collisions either always result in adsorption ($\frac1\tdi=0$) or never result in adsorption ($\frac1\tdA=0$) for simplicity.

\subsection{Coulumb drag}
\label{sec:Coulumbdrag}
For dust-ion drag when the ion kinetic energy is less than the dust surface potential, we assume a Coulumb drag model.\cite{drainePhysicsDustGrains1979,hopkinsResonantDragInstability2018}
\begin{equation}
\begin{split}
\tdi=&\htdi\left(\frac{k_B T_i}U\right)^2\frac1{\ln(\Lambda)}\left(1+a_C\left(\frac{|\ud-\ui|}\csi\right)^3\right)\\
\zeta_{si}=&1+2(\gamma_i-1)\Gamma-\frac{3(\gamma_i-1)}{2(1+a_C\hat w_i^3)}-\frac{1-(3-2\Gamma)(\gamma_i-1)}{2\ln{\Lambda}}\\
\zeta_{wi}=&-\frac{3a_C \hat w_i^3}{1+a_C \hat w_i^3}\\
\end{split}
\end{equation}
with definitions
\begin{equation}
\begin{split}
\htdi=&\left(\frac{\pi\gamma_i}{8}\right)^{\frac12}\frac{\brhod\Rd}{m_i n_{i0}{c_{si}}_0}\\
U=&\frac{e^2Z}\Rd\\
a_C=&\sqrt\frac{2\gamma_i^2}{9\pi}\\
\hat w_i=&\frac{u_{d0}-u_{i0}}{c_{si0}}
\end{split}
\end{equation}
where $\gamma_i$ is the ion polytropic index, $\ln(\Lambda)$ is the Coulumb logarithm, and $\Gamma$ is a temperature dependence parameter such that $Z\sim T^\Gamma$ (for our purposes $\Gamma=0$).

\subsection{Neutral-ion drag}
\label{sec:nidrag}
When it is necessary to calculate a value for the neutral-ion collision rate, we take $\nuni=\frac{\mi\ni}{\mi+\mn}\langle\sigma v\rangle_{mt}$, where $\langle\sigma v\rangle_{mt}$ is  the velocity-averaged momentum transfer cross-section for neutral-ion collisions. We use a formula given by Draine\cite{drainePhysicsInterstellarIntergalactic2011} for this:
\begin{equation}
\begin{split}
\langle\sigma v\rangle_{mt}=&1.21\cdot8.980\times10^{-10}\left(\frac{\alpha_N}{a_0^3}\right)^{1/2}\times\\
&\left(\frac{(\mi+\mn)m_p}{\mi\mn}\right)^{1/2}cm^3/s
\label{equ:sigmav}
\end{split}
\end{equation}
We specify the neutral polarizability $\alpha_N$ in units of $a_0^3$ where $a_0=\hbar/m_e e^2$.

\bibliography{bibliography}

\end{document}